\documentclass[preprint,12pt,authoryear]{elsarticle}

\usepackage{amssymb}
\usepackage{amsthm}
\usepackage{color}
\usepackage{hyperref}
\usepackage{color}
\usepackage{amsmath}
\usepackage{tabularx}
\usepackage{adjustbox}
\usepackage[T1]{fontenc}
\usepackage[utf8]{inputenc}

\newcommand{\aap}{    {Astron Astrophys }}
\newcommand{\aaps}{   {Astron Astrophys Suppl }}

\newcommand{\aj}{     {Astron J }} 
\newcommand{\an}{     {Astron Nachr }} 
\newcommand{\apj}{    {Astrophys J }}

\newcommand{\ajss}{   {Astrophys J, Suppl Ser }}
 
\newcommand{\araa}{   {Annu Rev Astron Astrophys }}

\newcommand{\mnras}{  {Mon Not Roy Astron Soc }}

\newcommand{\mcl}[1]{\multicolumn{2}{c}{#1}}
\newcommand{\mcc}[1]{\multicolumn{1}{c}{#1}}

\newcolumntype{R}{>{$}r<{$}}
\newcolumntype{L}{>{$}l<{$}}
\newcolumntype{A}{R@{${}\pm{}$}L}
\newcolumntype{E}{R@{${}-{}$}L}
\newcolumntype{B}{R@{${}\,/\,{}$}L}
\newcommand{\hpm}{$\,\pm\,$}

\journal{New Astronomy}

\begin{document}

\begin{frontmatter}



\title{A Dynamical Evolution Study of the Open Clusters: Berkeley 10, Berkeley 81, Berkeley 89 and Ruprecht 135}


\author[i1]{Hikmet \c{C}akmak\corref{cor1}} 
\cortext[cor1]{Corresponding author}
\ead{hcakmak@istanbul.edu.tr}

\author[i1]{Y\"uksel Karata\c{s}}

\affiliation[i1]{organization={\.Istanbul University Faculty of Science, Department of Astronomy and Space Sciences},
            addressline={\.Istanbul \"Universitesi Merkez Kamp\"us\"u}, 
            city={\.Istanbul},
            postcode={34116}, 
            state={T\"urkiye},
            country={TR}}

\begin{abstract}
By utilising Gaia EDR3 photometric/astrometric data, we studied the dynamical evolution from the obtained astrophysical, structural and dynamical parameters of the open clusters (OCs), Berkeley\,10 (Be~10), Berkeley\,81 (Be~81), Berkeley\,89 (Be~89), and Ruprecht\,135 (Ru~135). The Gaia EDR3 photometric distances from the isochrone fitting method are smaller than the ones of Gaia EDR2. The relaxation times of four OCs are smaller than their ages, in this regard, they are dynamically relaxed. Their steep overall mass function slopes mean that their low mass stars outnumber their massive ones. Their large $\tau$/relatively small $t_{rlx}$ values imply an advanced mass segregation. Therefore, they seem to have lost their low-mass stars much to the field. Be~89's outer parts indicate an expansion with time. However, Be~10 and Be~81 show the relatively shrinkage core/cluster radii due to dynamical evolution. Ru~135 (1.0~Gyr) may have a primordial origin, instead of shrinking in size and mass with time. Be~89's tidal radius is less than its cluster radius. This means that its member stars lie within its tidal radius, in the sense it is gravitationally bound to the cluster. For the rest OCs, the cluster members beyond their tidal radii are gravitationally unbound to the clusters, which are more influenced by the potential of the Galaxy.
\end{abstract}


	


\end{frontmatter}


\section{Introduction}\label{intro}
Cluster members of the OCs evolve dynamically due to internal (mass loss during stellar evolution, mass segregation, evaporation) and external perturbations such as encounters with the disk and Giant Molecular Clouds (GMCs), spiral arm shocks, and tidal stripping due to the galactic potential. The OCs lose a significant fraction of their primordial masses through via these processes \citep{Lamers2006} and \citep{Gieles2007}. As emphasized by \cite{Converse2011}, inner cores of star clusters contract, and thus their outer parts expand by transferring energy to the outer skirts. In the sense mass segregation drives the low mass stars to the outer parts of the clusters.  Most of the original mass are lost to the field because of dynamical effects along the several Myrs since their formation. 

From the analysis of Gaia EDR3 data of 389 OCs, \cite{Tarr2021} report that the cluster dimensions and mass segregation depend on cluster ages, in this regard, their core radii are on the average small for old OCs. They also state that mass segregation is more effective in old OCs. 

In this paper we study the dynamical evolution of Be~81, Ru~135, Be~10, and Be~89 from Gaia~EDR3 photometric/astrometric data. The first two are located in the inner galaxy and the other two are outside the solar circle. These old-aged OCs have been selected as representatives of first and second galactic quadrants in Fig.~1, where their positions relative to the spiral arms are displayed. Location criterion is important due to the survival rate of the OCs. As discussed by \cite{Bonatto2007b} and \cite{Gunes2017}, the majority of OCs with $age > 1$ Gyr lie outside the solar circle. The OCs inside the solar circle are scarce in the direction of galactic centre, due to the effects of strong absorption/crowding/dissolution by GMCs. \cite{Bonatto2007b} reported a deficiency of OCs older than 0.85 Gyr inside solar radius.  Within the framework of these issues, the astrophysical parameters (reddening, distance, age), structural parameters $(R_{RDP},~R_{core})$, tidal radii ($R_{t}$), overall masses, mass function slopes ($\chi$, MFs), time scales (relaxation time $t_{rlx}$, evolutionary parameter $\tau$, dissociation time $t_{dis}$) of these OCs have been derived. Our sample OCs were studied before in the literature with 2MASS JHK$_{s}$ and UBVRI photometries (Table~2). \cite{cantat2020} also provided their astrophysical parameters from Gaia~DR2 data. 

The organization of this paper is as follows. The separation of the cluster members is explained in Section~2. The determination of reddenings, distance moduli/distances, ages, and structural parameters of four OCs  is given in Section~3. Masses/mass function slopes, time scales (relaxation and dissolution times and evolutionary parameters), tidal radii, half-mass radii of four OCs are presented in Sections~4-5. Kinematics and orbital parameters are given in Section 6. A Discussion/Conclusion is presented in Section~7. 
\begin{figure}[!t]\label{fig-1}
	\centering{\includegraphics[width=0.6\columnwidth]{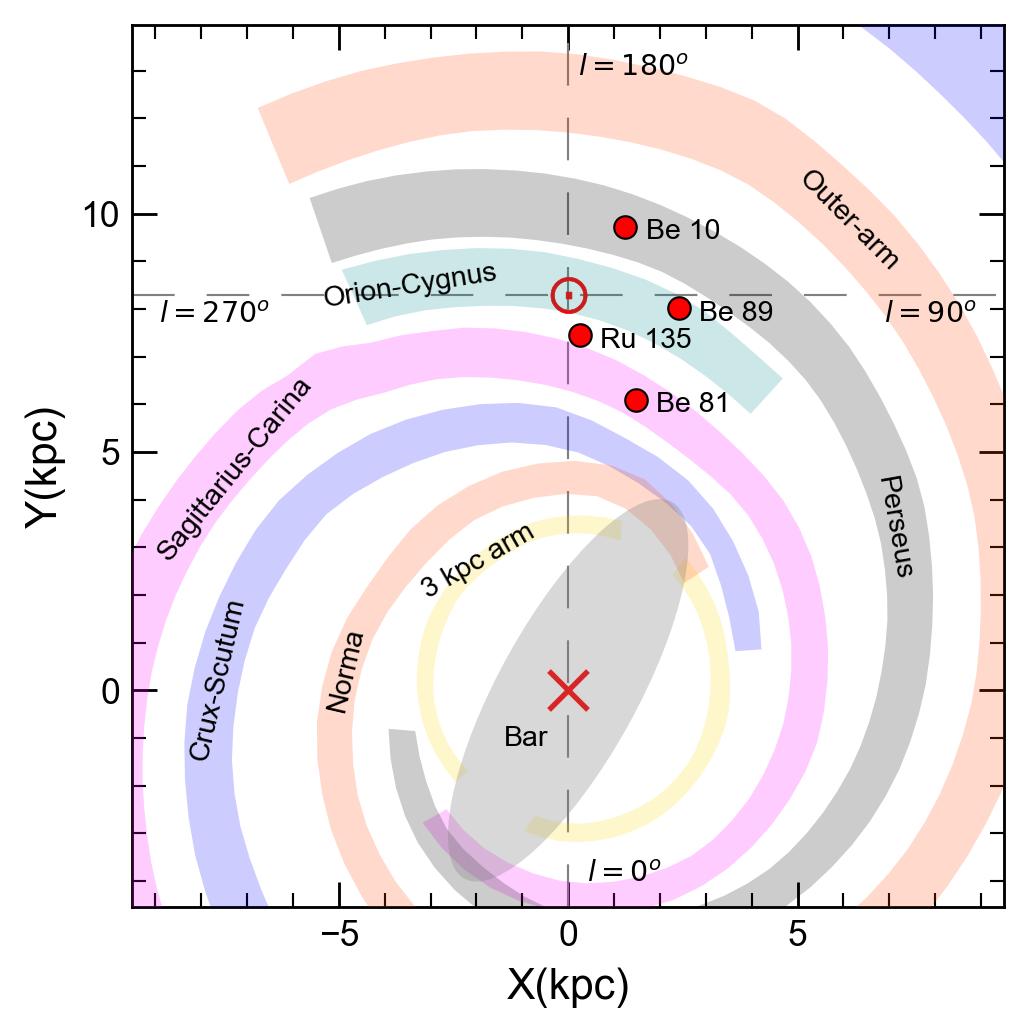}}
	\caption{Spatial distribution of four OCs  (filled red circles). The schematic projection of the Galaxy with its spiral arms is seen from the North pole. $(X,~Y)$~kpc show the galactosentric cartesian coordinates. The sun is at 8.2 kpc.}
\end{figure}

\section{Separation of the Cluster Members}
To separate the cluster members of our sample OCs from the field stars, their Gaia EDR3 astrometric/photometric data \citet{bro18} have been downloaded for a large-area (10 arcmin) from \begin{figure}[!t]\label{fig-2}
	\centering{
		\includegraphics[width=0.35\columnwidth]{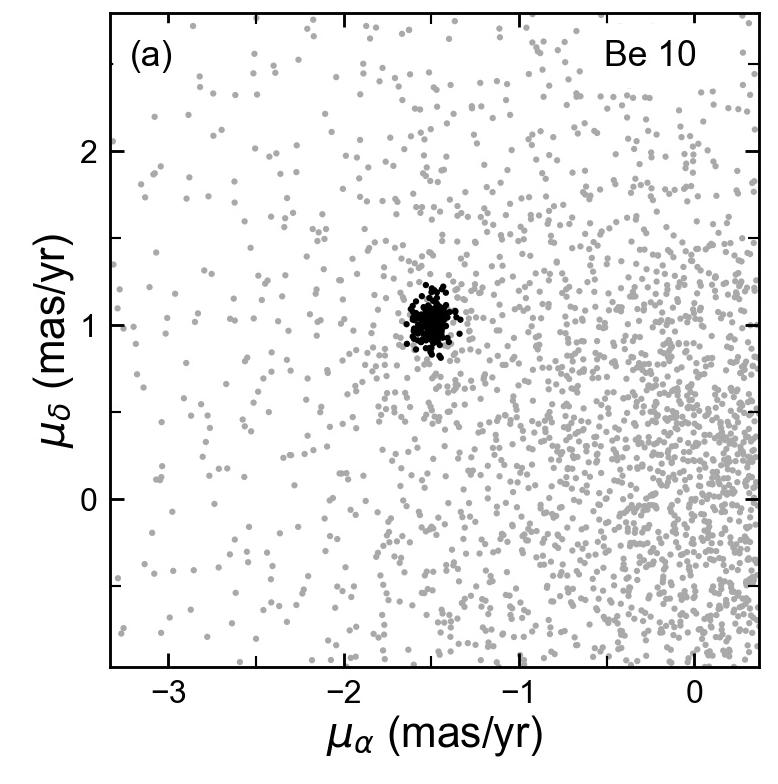}\hspace{5mm}
		\includegraphics[width=0.345\columnwidth]{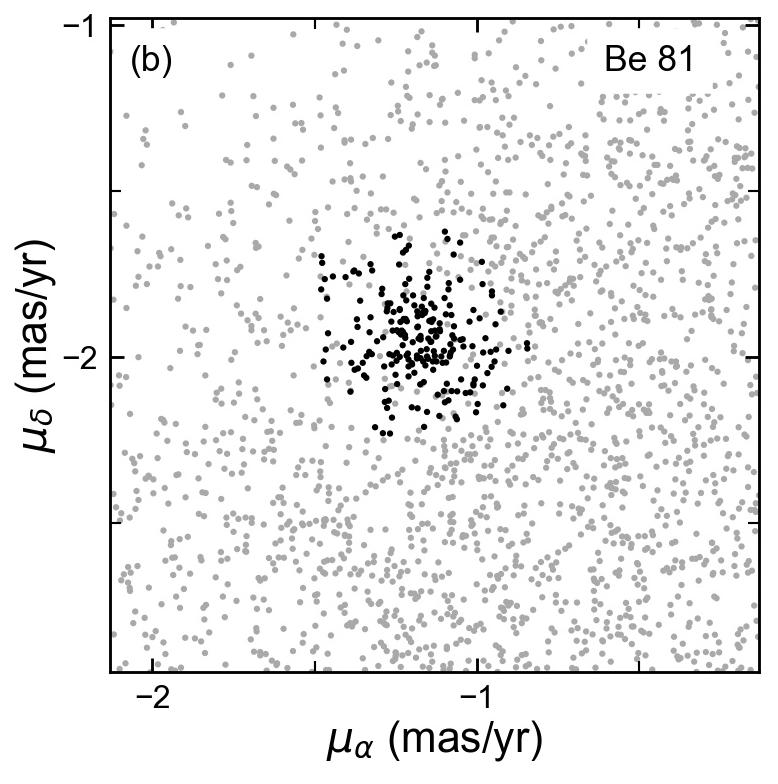}\vspace{3mm}
		\includegraphics[width=0.35\columnwidth]{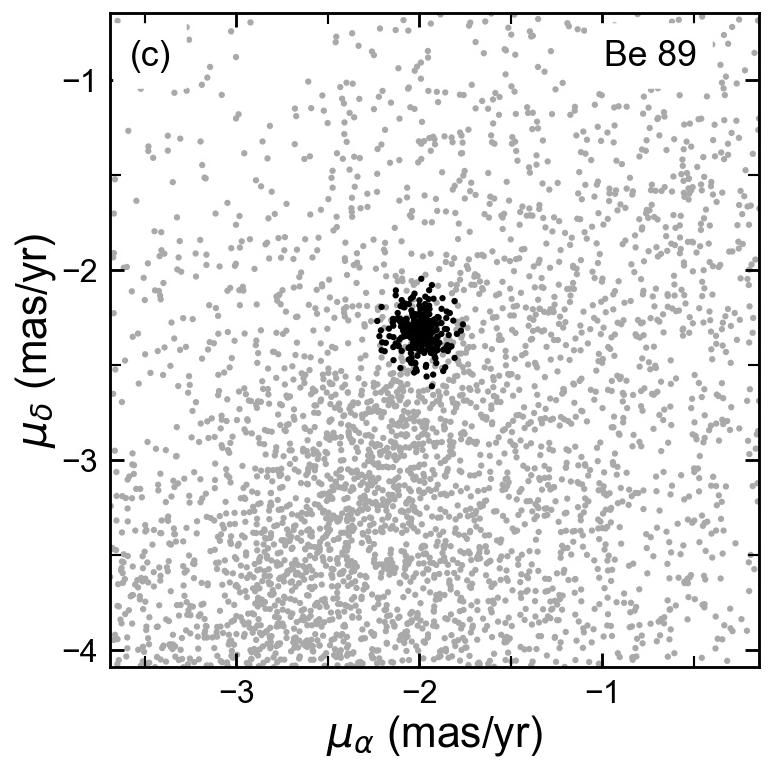}\hspace{5mm}
		\includegraphics[width=0.345\columnwidth]{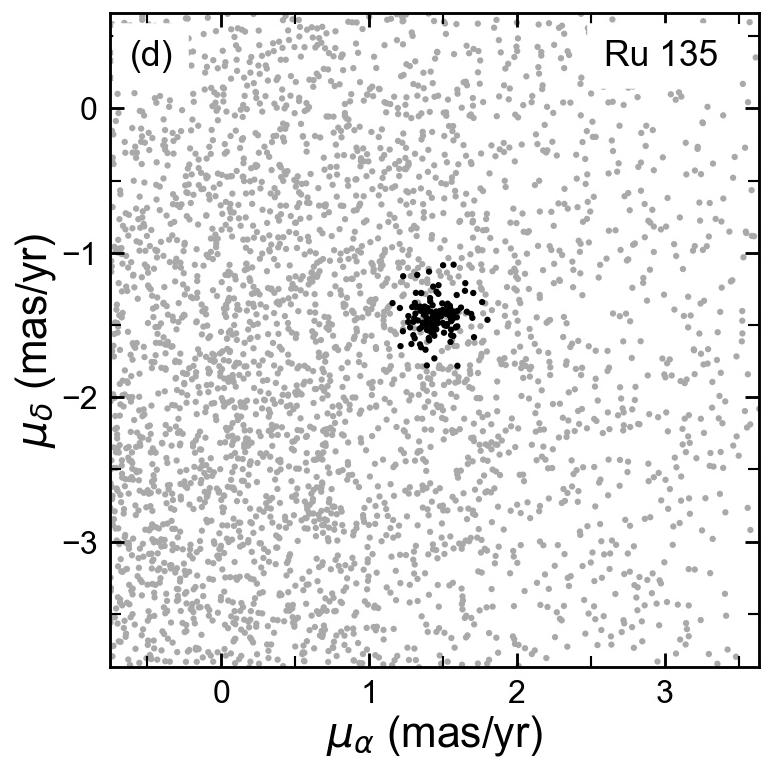}
	}
	\caption {$\mu_{\alpha}$ versus $\mu_{\delta}$ diagrams of four OCs. Filled dots denote the likely members, which are determined by using GMM. The grey points represent the field stars.}
\end{figure}
VizieR\footnote{http://vizier.u-strasbg.fr/viz-bin/VizieR?-source=II/246.}. On the ($\mu_{\alpha}$,~$\mu_{\delta}$) diagrams (Fig.~2), the potential cluster members show a more concentrated structure, whereas field stars have a more scattered distribution. We have applied  Gaussian Mixture Model (GMM) \citep{ped11} to the cluster stars. During GMM model trials, more than two suitable components were selected to clearly obtaion the central density in the vector point diagrams. Thus, the membership probabilities P(\%) of each cluster have been determined appropriately. The first significant rise in Fig.~3 is taken into account as the membership percentage (vertical dashed blue line). From Fig.~3, the probable members of four OCs seem to have greater than 85$\%$. The GMM \footnote{$P$ is defined $\Phi_c$ /$\Phi$.  Here $\Phi = \Phi_c + \Phi_f$ is the total probability distribution. \textit{c} and \textit{f} subscripts for cluster and field parameters, respectively. The used parameters for estimation of $\Phi_c$ and $\Phi_f$ are $\mu_{\alpha}$, $\mu_{\delta}$, $\varpi$, $\sigma_{\mu\alpha}$, $\sigma_{\mu\delta}$, $\sigma_\varpi$.} model considers that the distribution of proper motions of the stars in a cluster's region can be represented by two elliptical bivariate Gaussians. For the used parameters, see the papers of \citet{bal98}, \citet{wu02}, \citet{sar12} and \citet{dia18}.
\renewcommand{\tabcolsep}{1mm}
\renewcommand{\arraystretch}{1.6}
\begin{table*}[!t]\label{table-1}
	\centering
	\caption{Equatorial/Galactic coordinates, the median proper motion components, parallaxes/distances, and the numbers of the probable members of four OCs (top rows) and \citet{cantat2018,cantat2020} (bottom rows).}
	\begin{adjustbox}{width=1\textwidth}
		\begin{tabular}{lrrrrAAAAc}
			\hline
			Cluster &\mcc{$\alpha_{2000}(h\,m\,s)$}&\mcc{$\delta_{2000}(^{\circ}\,^{\prime}\,^{\prime\prime})$}&\mcc{$\ell(^{\circ})$}&\mcc{$b(^{\circ})$} & \mcl{$\mu_{\alpha}$ (mas/yr)} & \mcl{$\mu_{\delta}$ (mas/yr)} & \mcl{$\varpi$ (mas)} & \mcl{$d(kpc)$}  &  N\\  
			\hline
			Be\,10   &  3 39 30.9  &  66 29 08.7  &  138.61  &  8.88 & -1.50&0.03  &  1.01&0.06 & 0.39&0.06 & 2.45&0.36 & 197\\
			Be\,81   & 19 01 41.2  & -00 27 25.7  &   33.69  & -2.49 & -1.18&0.08  & -1.93&0.07 & 0.24&0.06 & 3.70&1.01 & 171\\
			Be\,89   & 20 24 24.9  &  46 02 03.1  &   83.13  &  4.84 & -1.98&0.06  & -2.32&0.06 & 0.28&0.06 & 3.35&0.64 & 221\\
			Ru\,135  & 17 57 59.4  & -11 39 59.8  &   16.38  &  6.27 &  1.44&0.07  & -1.46&0.05 & 0.89&0.06 & 1.11&0.07 & 119\\
			\hline\hline
			Be\,10   & & & & & -1.47&0.07  &  1.05&0.11 & 0.37&0.05 & 2.37&0.34 & 179\\
			Be\,81   & & & & & -1.20&0.16  & -1.85&0.16 & 0.25&0.09 & 3.31&1.17 & 127\\
			Be\,89   & & & & & -2.01&0.13  & -2.25&0.14 & 0.24&0.08 & 3.51&1.10 & 176\\ 
			Ru\,135  & & & & &  1.42&0.12  & -1.42&0.07 & 0.90&0.05 & 1.05&0.05 &  48\\
			\hline  
		\end{tabular}
	\end{adjustbox}
\end{table*}
\begin{figure}[!h]\label{fig-3}
	\centering{\includegraphics[width=0.50\columnwidth]{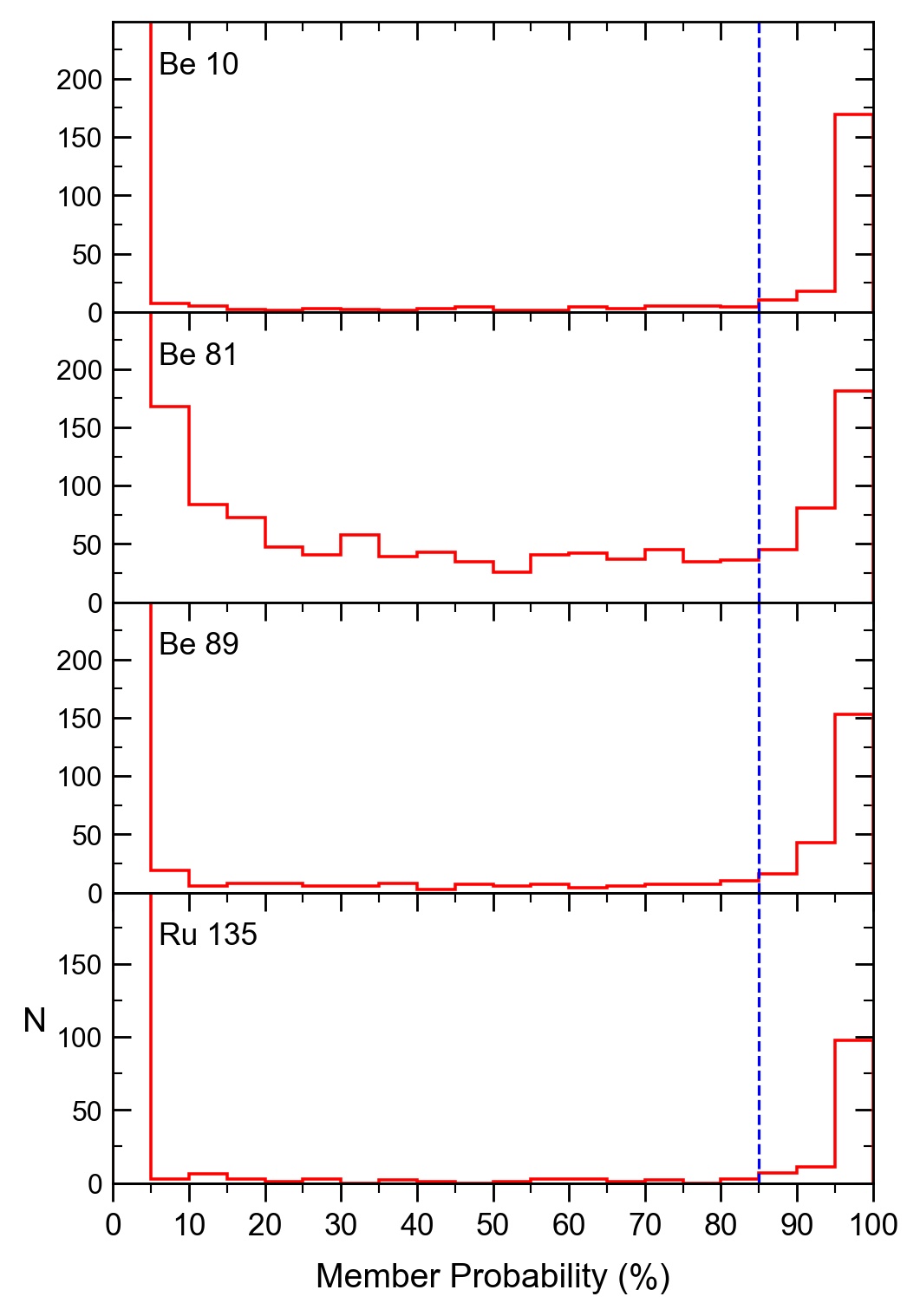}}
	\caption {The membership probability histogram P(\%). The vertical blue lines  indicate the selected lower limit of the probability.}
\end{figure}
The median proper motion components, the median parallaxes/distances, and the member numbers of four OCs are presented in Table 1. Instead of the inverse of Gaia EDR3 parallaxes, we averaged the parallaxes of cluster stars and then get the posterior probability density functions(PDFs) from those reduced errors, given by \citet{Bailer2018, Bailer2021}. For this, the global zero point is taken as $-$0.017 mas \citep{lin21}. Within the uncertainties, all the values are consistent with each other within the uncertainties, as compared to \citet{cantat2018,cantat2020} (bottom row). The estimated median equatorial coordinates from the likely members of our OCs and their corresponding galactic coordinates are listed in Table 1. These coordinates are similar to those of WEBDA \citep{Mermilliod1992}.

\section{Astrophysical and Structural Parameters}
\begin{figure}[t!]\label{fig-4}
	\centering{
		\includegraphics[width=0.32\columnwidth]{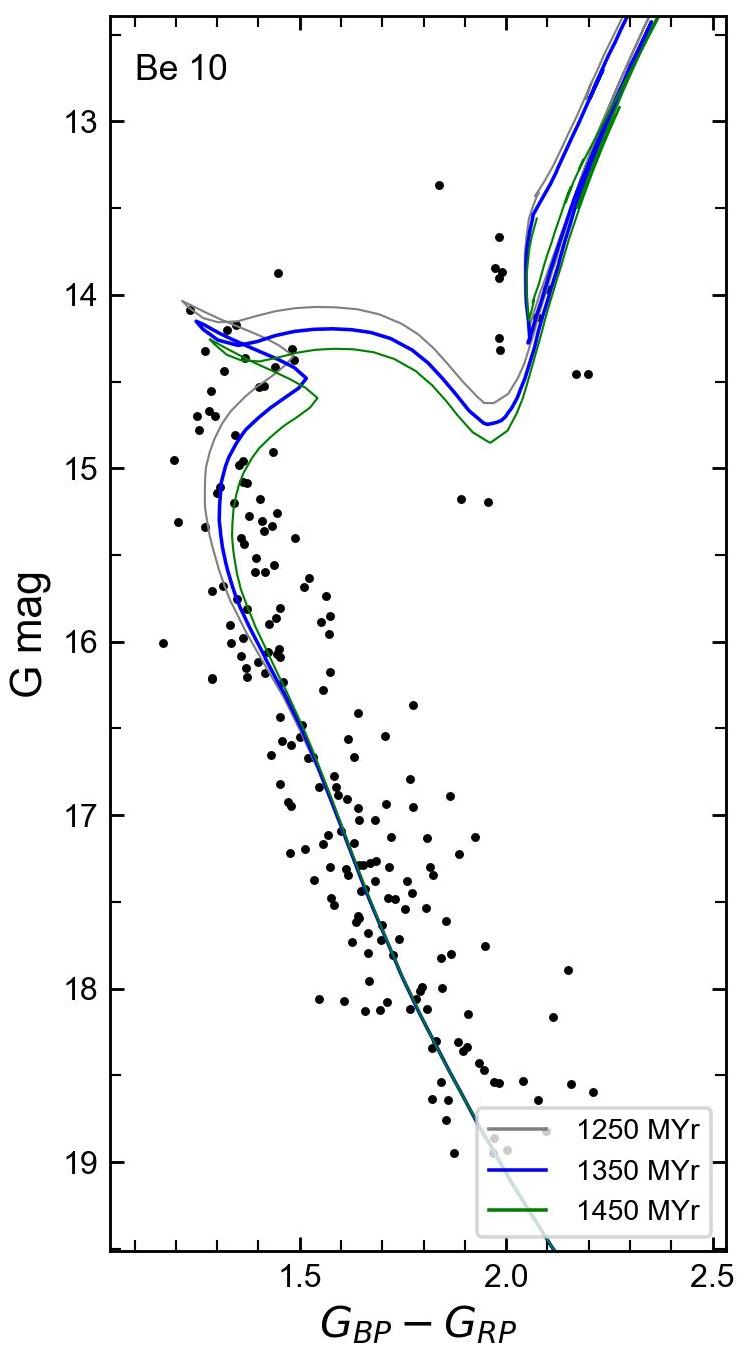}\hspace{5mm}
		\includegraphics[width=0.32\columnwidth]{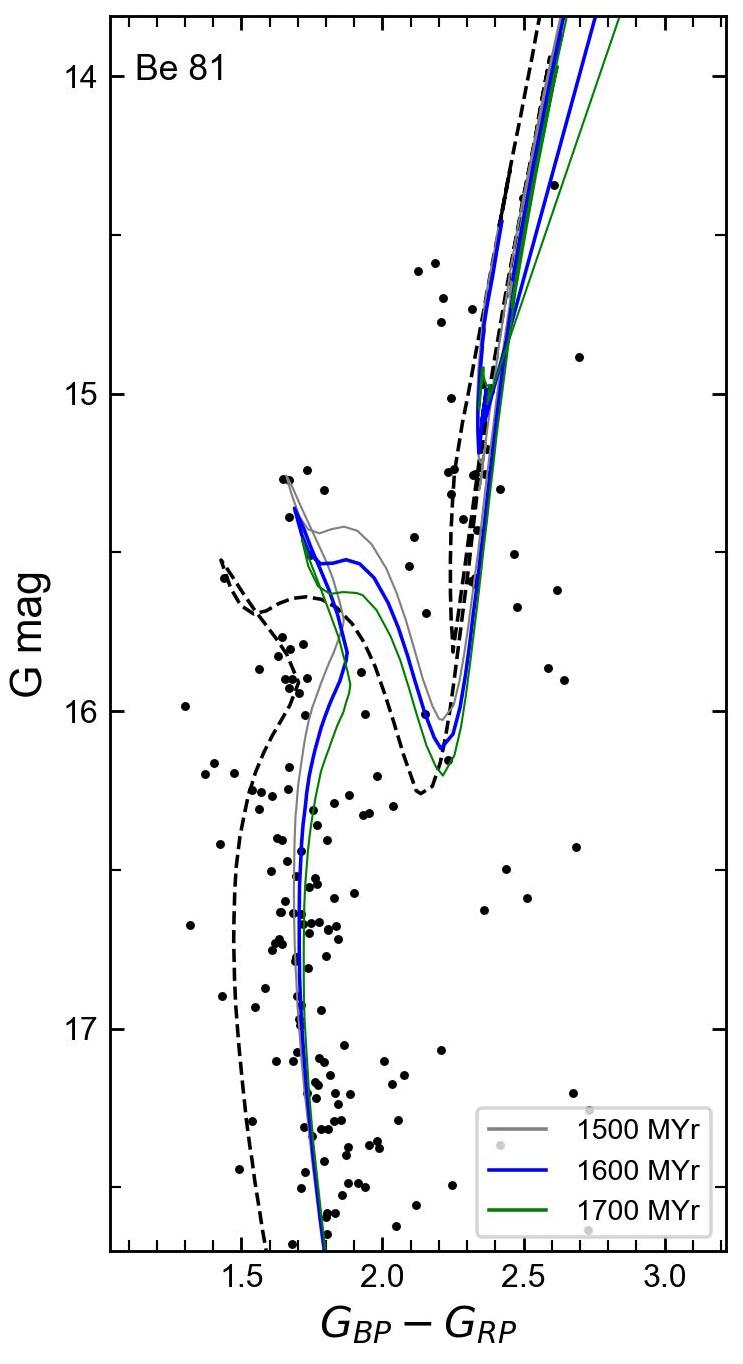}\\[2ex]
		\hspace{1mm}\includegraphics[width=0.33\columnwidth]{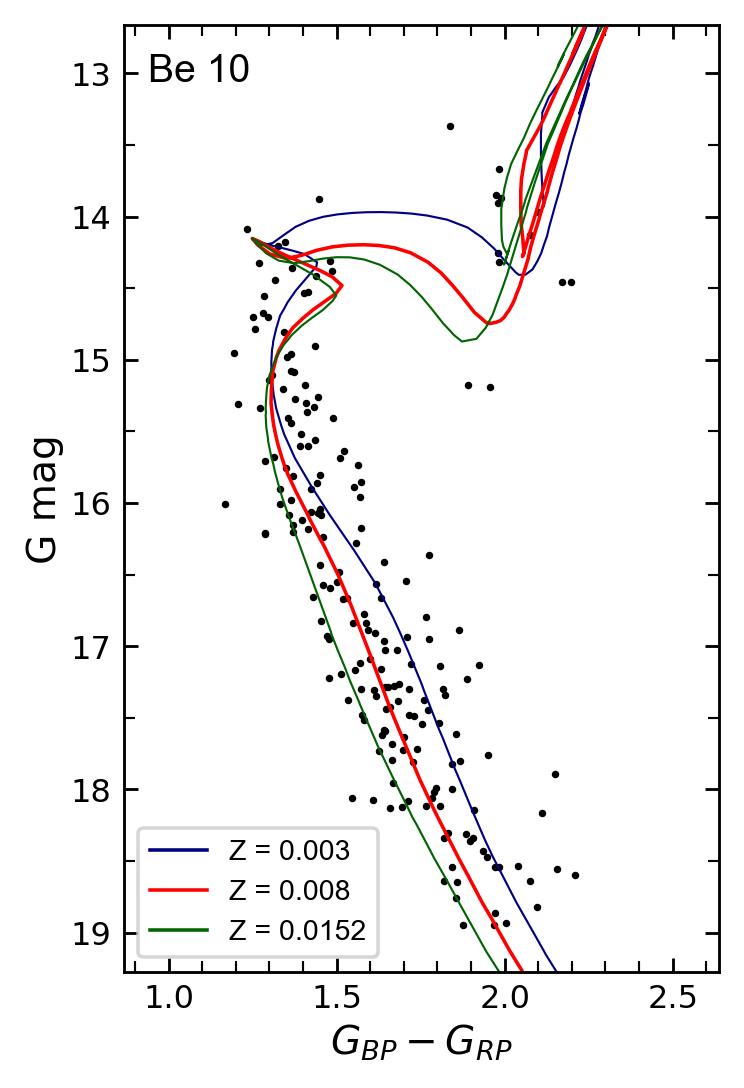}\hspace{4mm}
		\includegraphics[width=0.33\columnwidth]{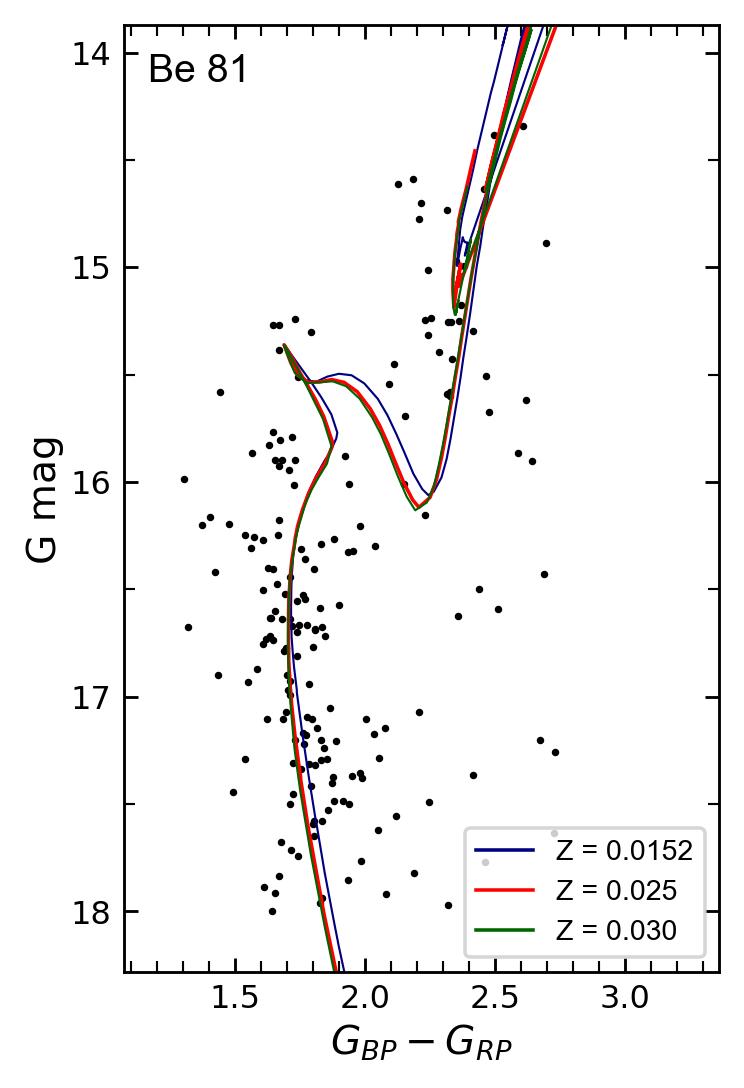}
	}
	\caption {The Gaia EDR3 CMDs of Be~10 and Be~81. The coloured solid lines represent the B12 isochrones in terms of Age (top panels) and $Z$-heavy element abundance (bottom panels).  The used ages and heavy element abundances for the fitted isochrones are indicated in the insets. The isochrones of $(Z=0.008,~1.35~Gyr)$ for Be~10 and of $(Z=0.025,~1.6~Gyr)$ for Be~81 provide satisfactory fits. For Be 81, the dashed isochrone shows the fit of \cite{cantat2020}.}
\end{figure}
The astrophysical parameters from the probable members of four OCs have been determined by fitting the observational $(G, G_{BP}-G_{RP})$ colour-magnitude diagrams (CMDs) to the  PARSEC isochrones of \citet{Bre12} (hereafter B12). \citet{Magrini2015} give the spectroscopic metal abundance of Be~81 as $[Fe/H] = +0.23\pm0.08$. This corresponds to heavy element abundance $Z = 0.025$. For the rest OCs, their spectroscopic metal abundances are not available in the literature. Therefore, several iterations by taking the isochrones of three heavy element abundances ($Z=0.008, 0.0152, 0.025$) together with an appropriate age are performed on the Gaia CMDs. The isochrones with $(Z, Age~(Myr)) =$ (0.008,~1350) for Be~10, (0.025,~1600) for Be~81, (0.0152,~3200) for Be~89, and (0.0152,~1000) for Ru~135, respectively provide satisfactory fits (Figs.~4-5). The PARSEC isochrones are shifted both vertically and horizontally on the Gaia CMDs to obtain the best fit to the observed main sequence, as well as the RG/RC (Red Giant/Red Clump) sequence, according to the interstellar reddening values of $E(G_{BP}-G_{RP})$ (Col.~2 of Table~2). Thus,  distance modulus $DM =(V_{0}-M_{V})$ and age are obtained. The distance moduli $(V_{0}-M_{V})$, distances d(pc) and ages (Myr) of four OCs are given in Cols.~5-7 of Table~2.  Their $E(B-V)$'s are converted from their $E(G_{BP}-G_{RP})$'s via  $E(B-V) = 0.775 E(G_{BP}-G_{RP})$ \citep{wang2019}.\clearpage
\begin{figure}[t!]\label{fig-5}
	\centering{
		\includegraphics[width=0.32\columnwidth]{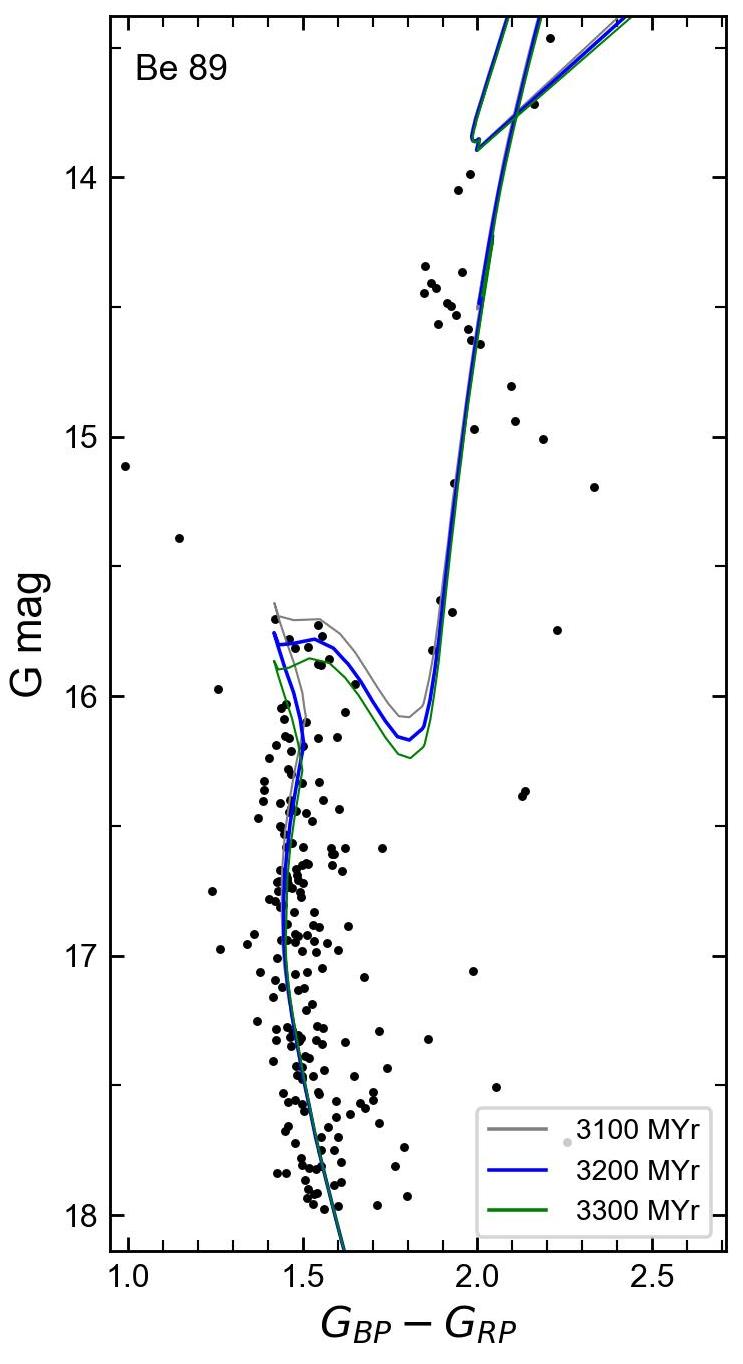}\hspace{5mm}
		\includegraphics[width=0.32\columnwidth]{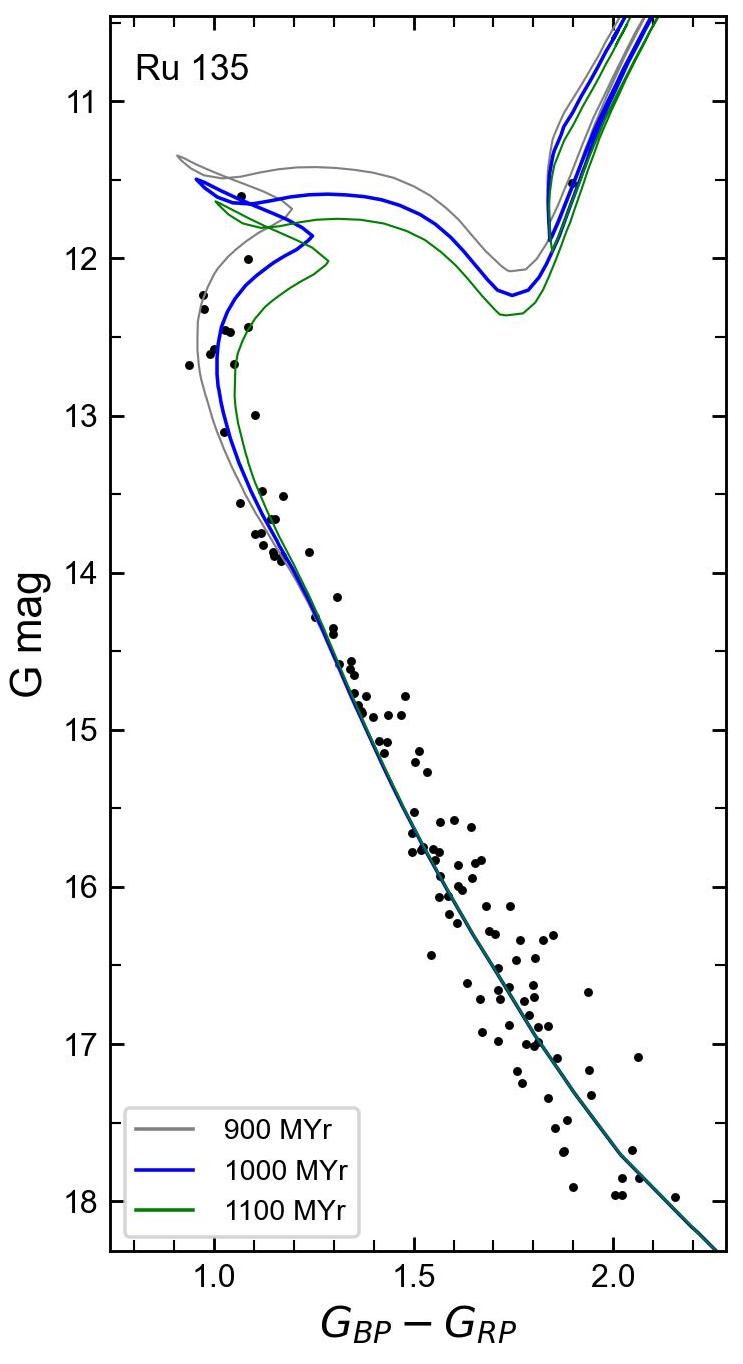}\\[3ex]
		\hspace{1mm}\includegraphics[width=0.33\columnwidth]{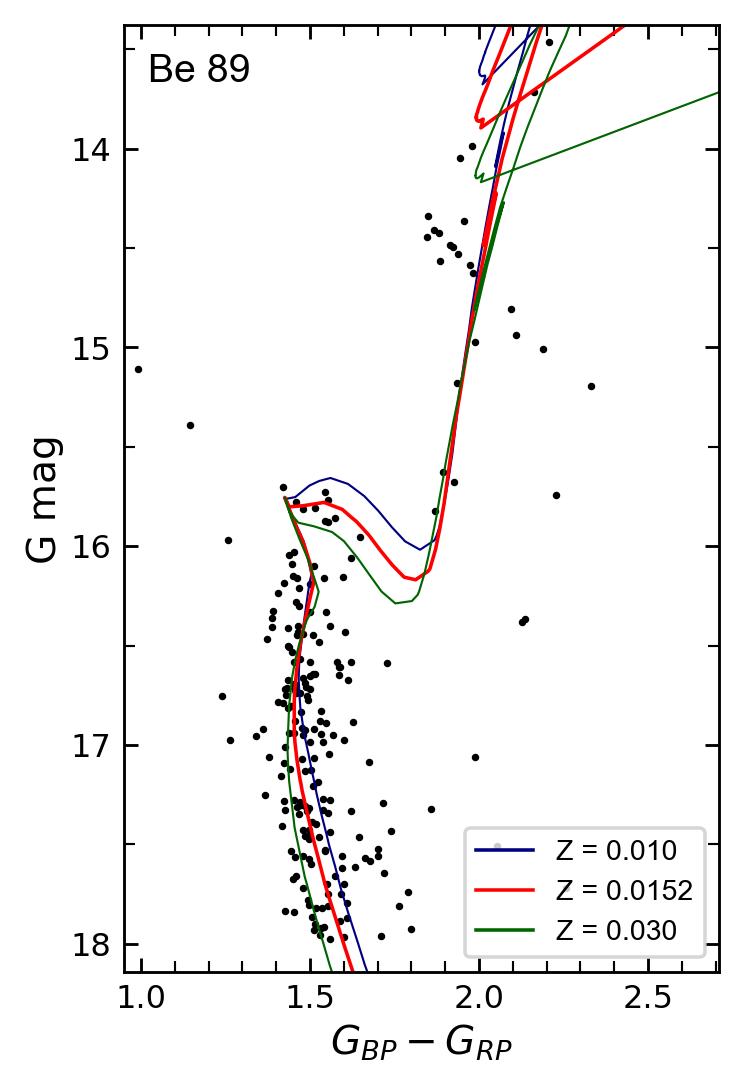}\hspace{4mm}
		\includegraphics[width=0.33\columnwidth]{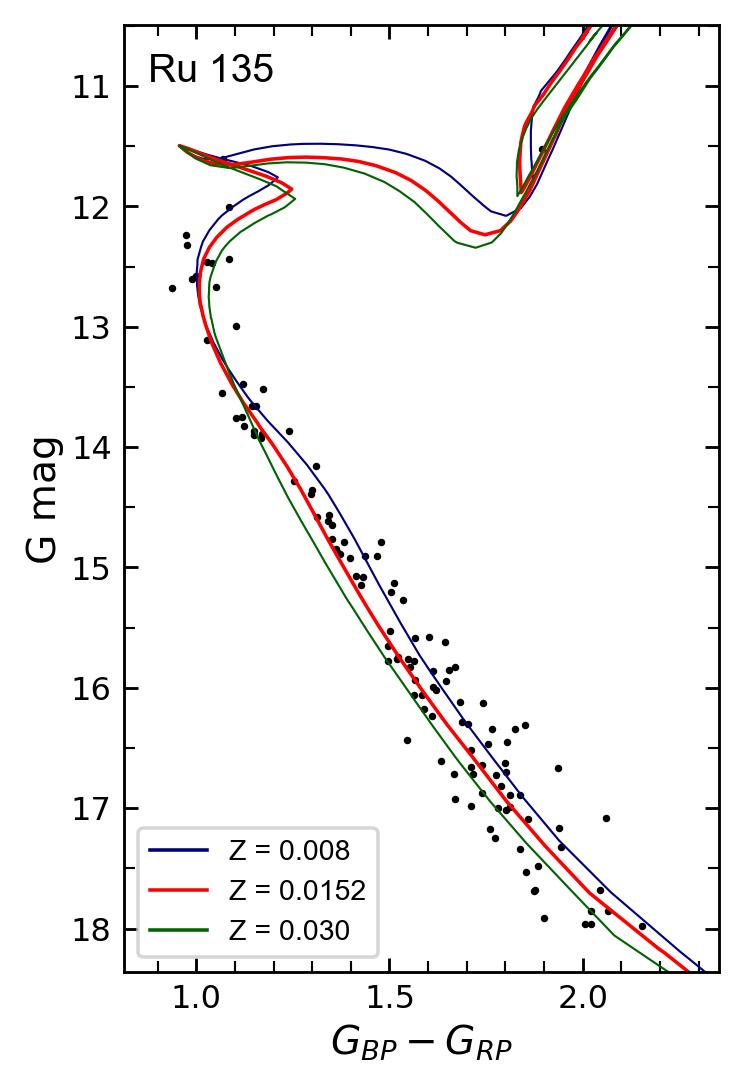}
	}
	\caption {The Gaia EDR3 CMDs of Be~89 and Ru~135. The symbols are the same as Fig.~4. The solar abundance isochrones, $(Z=0.0152,~3.2~Gyr)$ for Be~89 and $(Z=0.0152,~1.0~Gyr)$ for Ru~135 provide good fits to the cluster members.}
\end{figure}

\renewcommand{\tabcolsep}{1.1mm}
\renewcommand{\arraystretch}{1.2}
\begin{table*}[!t]\label{Table-2}
	\centering
	\caption{Comparison of the astrophysical parameters of four OCs  to the literature.}\vspace{1ex}
	\begin{adjustbox}{width=1\textwidth}
		\small
		\begin{tabular}{lccccccccll}
			\hline
			Cluster & $E(G_{BP}-G_{RP})$ & $E(B-V)$ &    Z    &$(V-M_{V})_{0}$ &    d(pc)     & Age (Myr)    &    Phot.  &    Isochrone           & Reference \\
			\hline
			Be~10   & 1.03$\pm$0.03 & 0.75$\pm$0.02 &  0.008  & 11.40$\pm$0.12 & 1909$\pm$96  & 1350$\pm$100 & Gaia EDR3 & Bressan et al.(2012)   & This work \\
				    &               &               &  solar  &      11.87     &     2365     &     1202     & Gaia DR2  & Bressan et al.(2012)   & (1) \\
				    &               &     0.80      &  solar  &      11.97     &     2201     &      646     &   2MASS   & Marigo et al.(2008)    & (2) \\
				    &               &     0.52      &  solar  &      11.92     &     1970     &     1000     &   2MASS   & Girardi et al.(2002)   & (3) \\
				    &               &     0.75      &  0.006  &      11.20     &     1700     &     1100     &   UBVRI   & Marigo et al.(2008)    & (4) \\
				    &               &     0.87      &  0.008  &      11.80     &     2290     &      631     &   UBVRI   & Girardi et al.(2002)   & (5) \\
				    &               &     0.71      &  solar  &      11.26     &     1790     &     1000     &     BV    & Bertelli et al. (1994) & (6) \\
			\hline         
			Be~81   & 1.16$\pm$0.03 & 0.85$\pm$0.02 &  0.025  & 12.12$\pm$0.10 & 2648$\pm$119 & 1600$\pm$100 & Gaia~EDR3 & Bressan et al.(2012)   & This work \\
			        &               &               &  solar  &      12.60     &     3313     &     1148     & Gaia DR2  & Bressan et al.(2012)   & (1) \\
			        &               &     1.06      &  solar  &      13.00     &     3401     &     1000     &   2MASS   & Marigo et al.(2008)    & (2) \\
			        &               &     0.90      &  solar  &      13.55     &     3601     &      891     &   2MASS   & Girardi et al.(2002)   & (3) \\
			        &               &     1.00      &  solar  &      15.60     &     3000     &     1000     &    BVI    & Bertelli et al.(1994)  & (7) \\
			        &               &     0.91      &  solar  &      12.40     &     3020     &      900     &    BVI    & Bressan et al. (1993)  & (8) \\
			\hline  
			Be~89   & 0.79$\pm$0.001& 0.58$\pm$0.001&  0.0152 & 11.93$\pm$0.11 & 2433$\pm$118 & 3200$\pm$100 & Gaia DR2  & Bressan et al.(2012)   & (1) \\
			        &               &     0.73      &  solar  &      12.50     &     2839     &     1585     &   2MASS   & Marigo et al.(2008)    & (2) \\
			        &               &     0.49      &  solar  &      12.99     &     3272     &     1585     &   2MASS   & Girardi et al.(2002)   & (3) \\
			        &               &     0.60      &  0.008  &      11.90     &     2400     &     3600     &   UBVRI   & Marigo et al.(2008)    & (4) \\
			        &               &     1.03      &  solar  &      13.30     &     3005     &      850     &   2MASS   & Bonatto et al.(2004)   & (9) \\
			        &               &     1.05      &  solar  &      14.80     &     2040     &     1122     &    BVI    & Girardi et al.(2000)   & (10) \\
			\hline                  
			Ru~135  & 0.74$\pm$0.05 & 0.54$\pm$0.04 &  0.0152 & 9.76$\pm$0.06  &  897$\pm$60  & 1000$\pm$100 & Gaia~EDR3 & Bressan et al.(2012)   & This work \\
			        &               &               &   solar &      10.12     &     1054     &      646     & Gaia DR2  & Bressan et al.(2012)   & (1) \\
			        &               &     0.87      &   solar &      13.00     &     3497     &     3162     &   2MASS   & Marigo et al.(2008)    & (2) \\
			        &               &     1.14      &   solar &      12.66     &     2172     &      562     &   2MASS   & Girardi et al.(2002)   & (3) \\
			        &               &     0.63      &   0.004 &       9.50     &      750     &     4000     &   UBVRI   & Marigo et al.(2008)    & (4) \\
			        &               &     1.10      &   solar &      11.33     &     1850     &      501     &   2MASS   & Bonatto et al.(2004)   & (11) \\
			\hline
		\end{tabular}%
	\end{adjustbox}
	\\[1ex] 
	\footnotesize \parbox{\textwidth}{References of the last column} \\[1ex]
	\renewcommand{\tabcolsep}{1mm}
	\renewcommand{\arraystretch}{0.9}
	\begin{tabular}{r l m{1cm} r l}
		 1 & Cantat-Gaudin  et al. (2018, 2020)  & &  7 & Sagar \& Griffiths (1998) \\
		 2 & Kharchenko  et al. (2013)           & &  8 & Donati et al. (2014)      \\ 
		 3 & Bukowiecki et al. (2011)            & &  9 & Tadross (2008a)           \\ 
		 4 & Akkaya et al. (2010)                & & 10 & Subramaniam et al. (2010) \\ 
		 5 & Lata et al. (2004)                  & & 11 & Tadross (2008b)           \\
		 6 & Maciejewski \& Niedzielski (2007)   & & & \\
	\end{tabular}
\end{table*} 

The stellar radial density profiles (RDPs) of four OCs are constructed from the Gaia EDR3 photometric/astrometric data for cluster members and the field within $10'.0$ down to $G =19$ mag. From the RDPs, their structural parameters are derived. The fitted RDPs of four OCs to the relation of \citet{King1966} are given in Fig.~6. For this, the three-parameter function \citep{King1966},
\begin{equation*}
	\sigma(R) = \sigma_{bg} + \frac{\sigma_0}{1+(\frac{R}{R_{core}})^2}
\end{equation*}
is considered. For this equation, $\sigma_{bg}$ is the residual background density. $\sigma_0$ and $R_{core}$ are the central \begin{figure}[t!]\label{fig-6}
	\centering{
		\includegraphics[width=0.385\columnwidth]{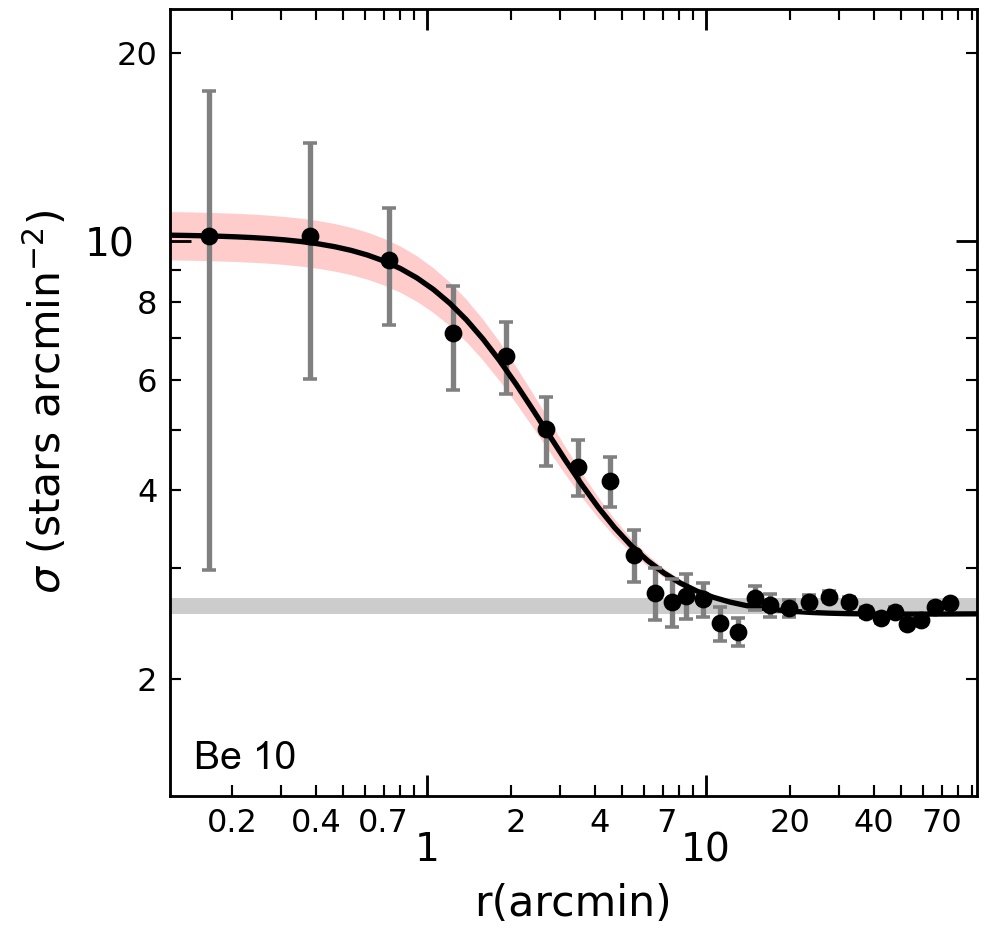}\hspace{5mm}
		\includegraphics[width=0.385\columnwidth]{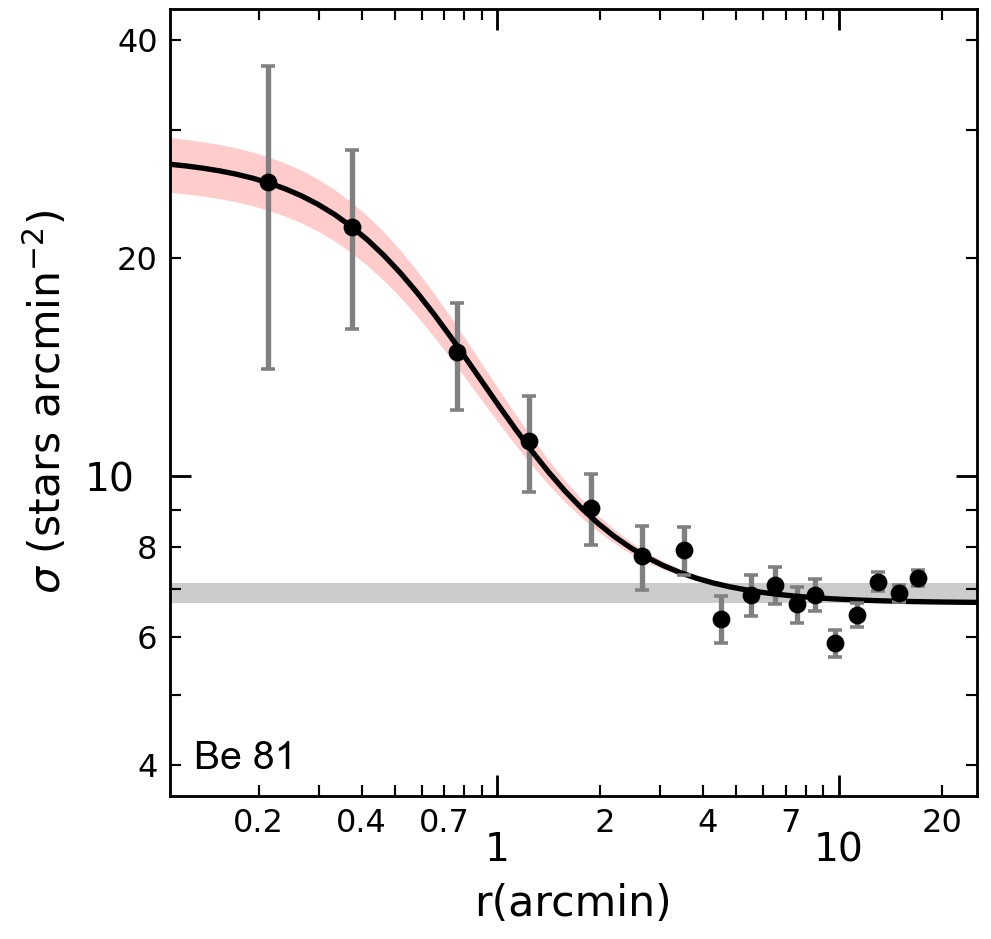}\vspace{3mm}
		\includegraphics[width=0.39\columnwidth]{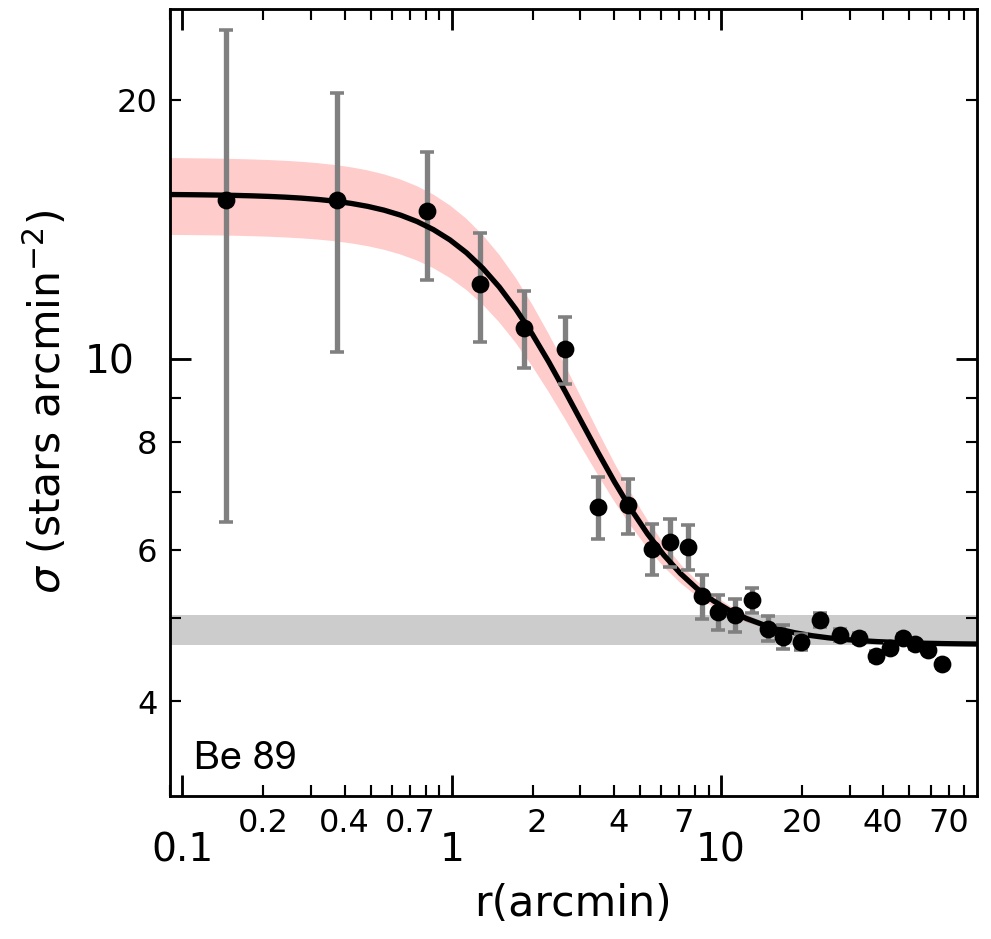}\hspace{5mm}
		\includegraphics[width=0.39\columnwidth]{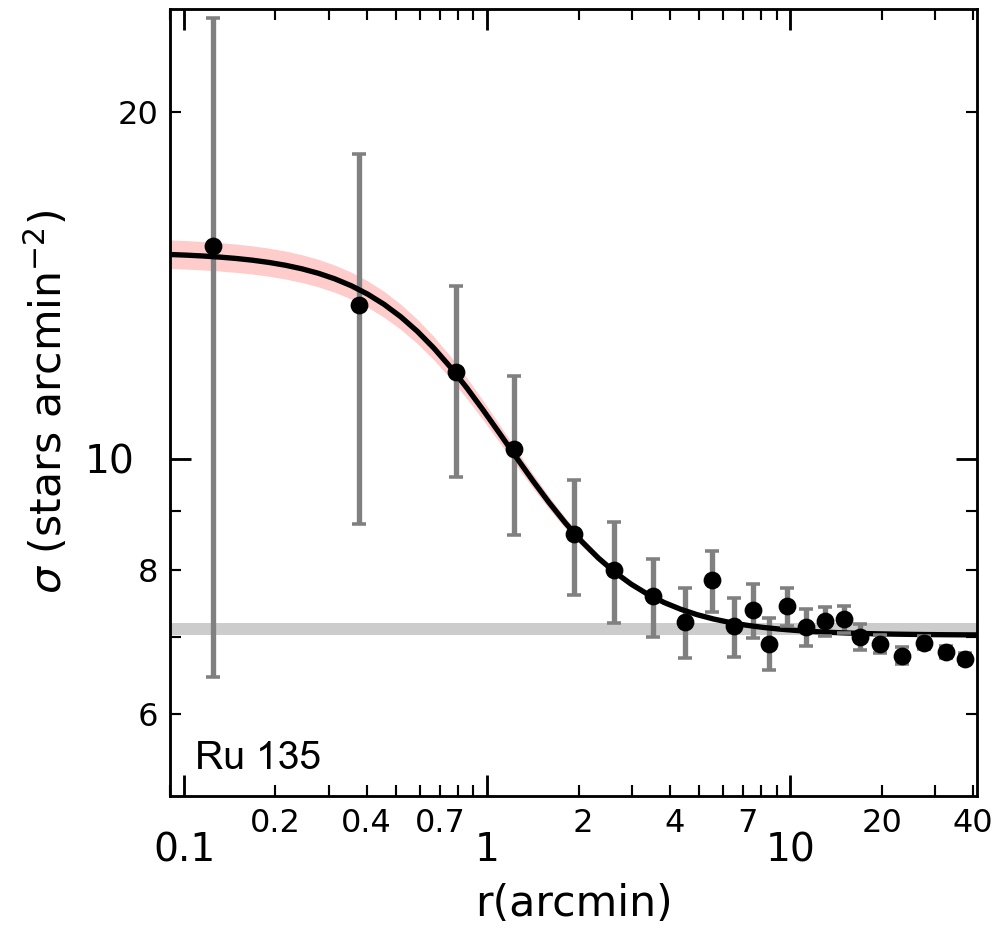}
	}
	\caption {The radial density profiles of four OCs. Solid line shows the best fit King profile and the horizontal gray bar denotes the stellar background level measured in the comparison field. The $1\sigma$ King fit uncertainty is shown by the red shaded domain.}	
\end{figure}
density of stars and the core radius, respectively. Thus, the $R_{core}$, $\sigma_{bg}$, and  $\sigma_0$ of four OCs have been derived by fitting King profile to the observational RDPs (Fig.~6). The meanings of solid line, horizontal bar and the shaded area are mentioned in caption of Fig.~6. 
\renewcommand{\tabcolsep}{4.9mm}
\renewcommand{\arraystretch}{1.3}
\begin{table*}[h!]\label{table-2}
	\begin{center}
		\caption{Structural parameters. Row 1: arcmin to parsec scale. The symbol $*\,\prime^{-2}$ in rows ~2--3 denotes $stars~arcmin^{-2}$. $\delta_{c}$ in row 4 means the density contrast parameter. Core and cluster radii are listed in rows.~5--7 in units of arcmin and pc, respectively. CC (last row) means the correlation coefficient. }\vspace{-1.5ex}
		\begin{adjustbox}{width=1\textwidth}
		\begin{tabular}{lAAAA}
			\hline
			Parameters                      & \mcl{Be\,10} & \mcl{Be\,81} & \mcl{Be\,89} & \mcl{Ru\,135}\\
			\hline
			$(1')\,(pc)$                    &  \mcl{0.58}  &  \mcl{0.67}  &  \mcl{0.87}  & \mcl{0.26}\\
			$\sigma_{0}\,(*\,\prime^{-2})$  &   8.32&0.98  &  20.81&2.76  &  10.90&0.75  &  7.02&0.09\\
			$\sigma_{bg}\,(*\,\prime^{-2})$ &   2.53&0.08  &   6.69&0.24  &   4.66&0.13  &  8.11&0.82\\ 
			$ \delta_{c}$                   &   4.28&0.52  &   4.11&0.56  &   3.34&0.25  &  2.17&1.05\\ 
			$R_{core}\,(\prime)$            &   1.90&0.15  &   0.63&0.03  &   2.22&0.20  &  0.96&0.08\\ 
			$R_{RDP}\,(\prime)$             &  11.23&0.43  &   4.51&0.29  &  13.02&0.57  &  6.51&0.30\\  
			$R_{core}\,(pc)$                &   1.10&0.09  &   0.42&0.02  &   1.93&0.17  &  0.25&0.02\\ 
			$R_{RDP}\,(pc)$                 &   6.50&0.25  &   3.02&0.19  &  11.38&0.50  &  1.69&0.08\\
			CC                              & \mcl{0.991}  & \mcl{0.995}  & \mcl{0.988}  & \mcl{0.989}\\
			\hline
			\citet{Bukowiecki2011} \\
			\hline
			$R_{core}\,(pc)$                &   1.36&0.18  &   0.63&0.08  &   0.95&0.12  &  0.45&0.07\\ 
			$R_{RDP}\,(pc)$                 &   7.96&1.30  &   5.88&0.94  &   8.23&1.18  &  3.42&0.62 \\
			\hline
			\citet{Kharchenko2013} \\
			\hline
			$R_{core}\,(pc)$                &  \mcl{0.80}  &  \mcl{1.28}  &  \mcl{1.67}  & \mcl{1.71}\\ 
			\hline
			\citet{Maciejewski2007} \\
			\hline
			$R_{core}\,(pc)$        &0.70&0.25  &\mcl{} &\mcl{} & \mcl{} \\ 
			$R_{RDP}\,(pc)$         &4.30&1.50  &\mcl{} &\mcl{} & \mcl{}  \\
				\hline
			\citet{sag1998} \\
			\hline
			$R_{RDP}\,(pc)$         &  \mcl{}   &  \mcl{2.35}  &  \mcl{}  &    \mcl{} \\
			\hline
		\end{tabular}
	\end{adjustbox}
	\end{center}
\end{table*}

Their cluster radii $(R_{RDP})$ by comparing the RDP level with background and by measuring the distance from the centre \citep{Bonatto2007a} are determined. The $R_{RDP}$ is taken as an observational truncation radius, whose value depends both on the radial distribution of member stars and the field density.  The large uncertainties within  $R < 1'$ in the RDPs are due to their low star contents in their central parts. The density contrast parameter ($\delta_{c}$) to estimate the compactness of the clusters is obtained from the relation,
\begin{equation*}
	\delta_{c} = 1 +\frac{\sigma_0}{\sigma_{bg}}
\end{equation*}
As emphasized by \cite{Bica2005}, the $\delta_{c}$ values of the OCs depend on the Galactic latitude than the cluster mass, accordingly, the populous and high Galactic latitude OCs have large density contrast parameters, i.e.: NGC~188 $(b= +22.^{\circ}39, \sim 3800 M_{\odot})$ and M~67 $(b= +31.^{\circ}89, \sim 990 M_{\odot})$. Different Galactic latitudes have different properties and also suffer from differences in density contrast, since low-Galactic latitude OCs tend to be projected against denser fore/backgrounds. As compared to the range  $(7 \le \delta_{c} \le 23)$, which is found from compact star clusters by \cite{Bonatto2009}, our $\delta_{c}$ values are lower. This indicates that our OCs with low-latitude are sparse.  These structural parameters and their meanings together with the literature values are listed in rows~1-9 of Table~3. 

As compared to literature values (bottom rows of Table 3), for the $R_{core}$(pc)'s of Be~10, Be~81 and Ru~135, there is in a reasonable agreement with the ones of \cite{Bukowiecki2011} within the uncertainties. Be~89's $R_{core}$ value is somewhat larger than \cite{Bukowiecki2011}.  \cite{Kharchenko2013} give a large $R_{core}$ (pc) value for Be~81 and Ru~135. For Be~10 and Be~89, their $R_{core}$ (pc) values are in good agreement with our findings within the error bars. Our $R_{RDP}$ values for Be~89 and Ru~135 are quite larger than the 8.23 pc value of \cite{Bukowiecki2011}. Within the error limits, our $(R_{core},~R_{RDP})$ values for Be~10 are in consistent with the ones of  \cite{Maciejewski2007}. For Be~81,  \cite{sag1998} give a slightly small  $R_{RDP}$ (pc) value. 

\section{Mass and Mass Function Slope}
The masses of the main sequence members of four OCs are determined from the B12-isochrones. $G$ magnitudes of main sequence members have been converted into the absolute magnitudes from their reddenings and distance moduli  (Cols.~2 and 5 of Table~2).  Mass function (MF) is defined as the distribution of masses of cluster stars per unit volume. The mass functions in terms of 0.05 bin sizes are shown in Fig.~7. The MF slope is fitted to a power-law given by,\\
\begin{equation*}
	\log\frac{dN}{dM}=-(1+x)\log(M) + \text{constant}
\end{equation*}\\
Where $dN$ is the probable cluster members in a mass bin $dM$ with central mass $M$ and $x$ is mass function slope. Since Gaia data ($G$ mag) is not complete below $G=19$ mag \citep{Arenou2018},  stars brighter than this limit are taken, which correspond to stars more massive than 1 $M_{\odot}$. The obtained MF slopes are $x=1.46\pm0.60$, $2.88\pm0.93$, $1.24\pm0.90$ and $1.99\pm0.44$ for Be~10, Be~81, Be~89, and Ru~135, respectively. The MFs of Be~10 and Be~89 are consistent with $\chi=1.3\pm0.7$ of \cite{Kroupa2001} with the uncertainties. Total masses are obtained from the MF slopes of four OCs. The mass range, MF slope,  total mass, mean mass, and member number are listed in Cols.~2-6 of Table~4. In order to see the mass-segregation effects of four OCs, the normalized cumulative radial stellar distributions of member stars for three mass ranges against cluster radius (arcmin) have been displayed in Fig.~8. For Be~10, bright stars (high mass ones) slightly outnumber low mass/fainter counterparts throughout the inner parts. Towards its outer parts, low-mass \renewcommand{\tabcolsep}{1.9mm}
\renewcommand{\arraystretch}{1.2}
\begin{table*}[t!]\label{Table-4}
	\begin{center}
		\caption{The mass information of four OCs calculated from their mass functions.}\vspace{1ex}
		\begin{tabular}{lcAAAA}
			\hline
			Cluster   & Mass range & \mcl{MF slope} & \mcl{Total mass} & \mcl{Mean Mass} & \mcl{Members} \\[-1.0ex]
			&  $M_{\odot}$ & \mcl{$\chi$}  & \mcl{$M_{\odot}$} & \mcl{$M_{\odot}$} &  \mcl{N}  \\
			\hline
			Be\,10  & 0.75 $-$ 1.13 & 1.46&0.60 & 109.6&13.1  & 0.90&0.15 & 122&14 \\
			Be\,81  & 0.91 $-$ 1.21 & 2.88&0.93 & 134.5&14.9  & 1.03&0.16 & 131&14 \\
			Be\,89  & 0.80 $-$ 1.04 & 1.24&0.90 & 154.7&17.5  & 0.90&0.14 & 171&19 \\
			Ru\,135 & 0.69 $-$ 1.51 & 1.99&0.44 &  98.5&16.8  & 0.90&0.20 & 109&16 \\
			\hline
		\end{tabular}
	\end{center}
\end{table*}\begin{figure}[!h]\label{fig-7}
	\centering{
		\includegraphics[width=0.39\columnwidth]{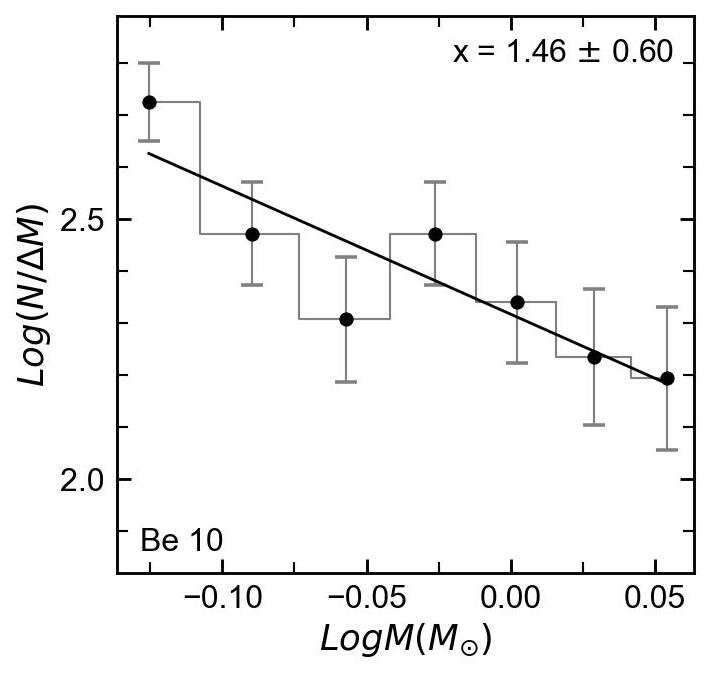}\hspace{5mm}
		\includegraphics[width=0.385\columnwidth]{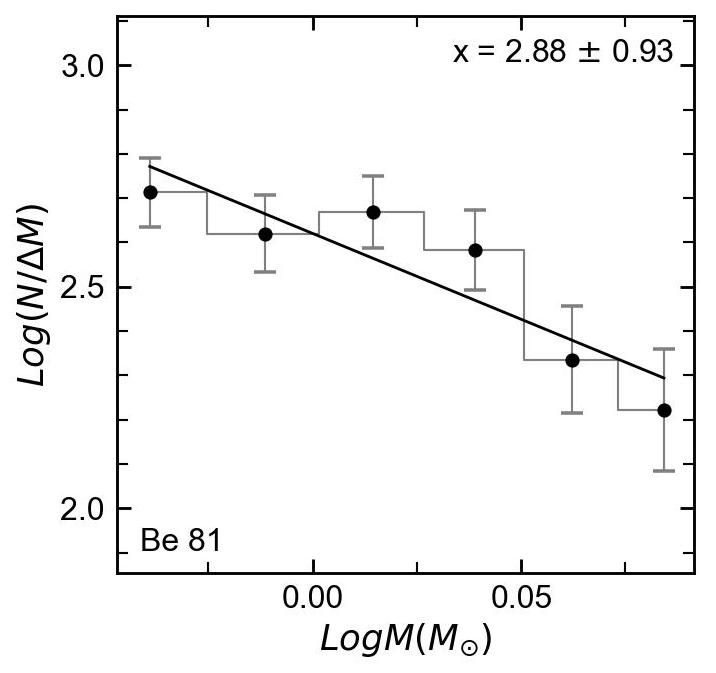}\vspace{3mm}
		\includegraphics[width=0.39\columnwidth]{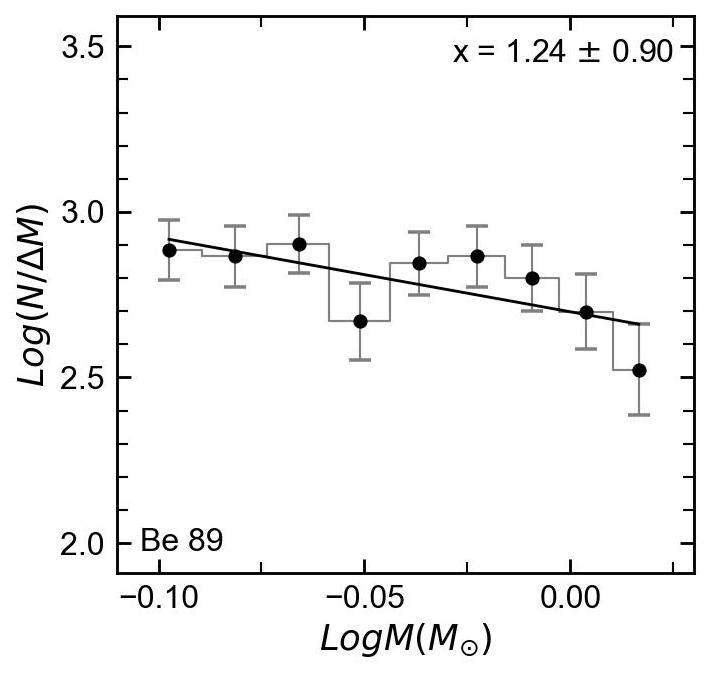}\hspace{5mm}
		\includegraphics[width=0.395\columnwidth]{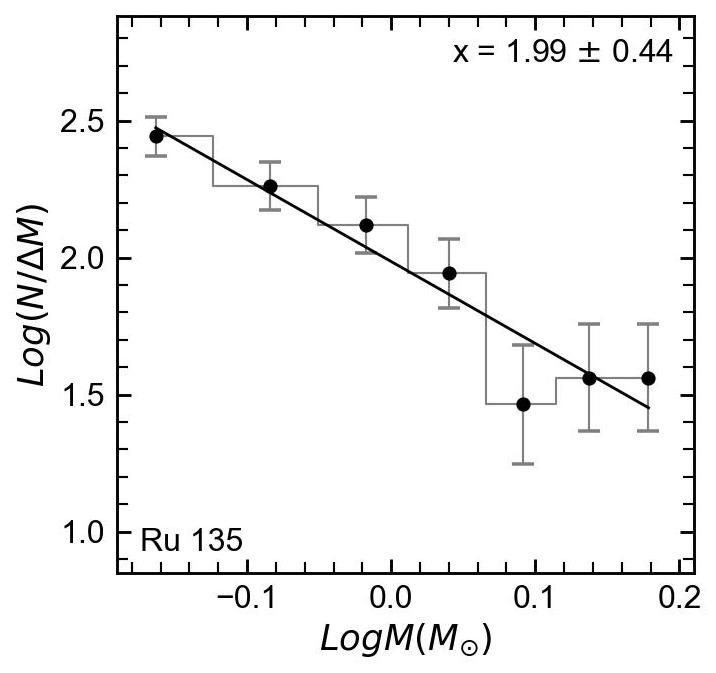}
	}
	\caption {The mass functions of four OCs. The vertical error bars are from $\frac{1}{\sqrt{N}}$.}
\end{figure}stars dominate. For Be~81 and Be~89, high mass stars outnumber low mass counterparts throughout the entire cluster radius.  For Ru~135'core region, high mass stars slightly outnumber low-mass stars. For $1 < r(arcmin) < 4$, intermediate mass stars seem to be dominant over low mass ones. Towards its skirts, high mass stars slightly outnumber the low mass ones.  During the lifetime of the OCs, encounters between its member stars lead to an increased degree of energy equipartition. Thus, the high  mass stars sink towards the cluster center and transfer their kinetic energy to the low-mass stars, thus leading to mass segregation.
\begin{figure}[!t]\label{fig-8}
	\centering{	\includegraphics[width=0.39\columnwidth]{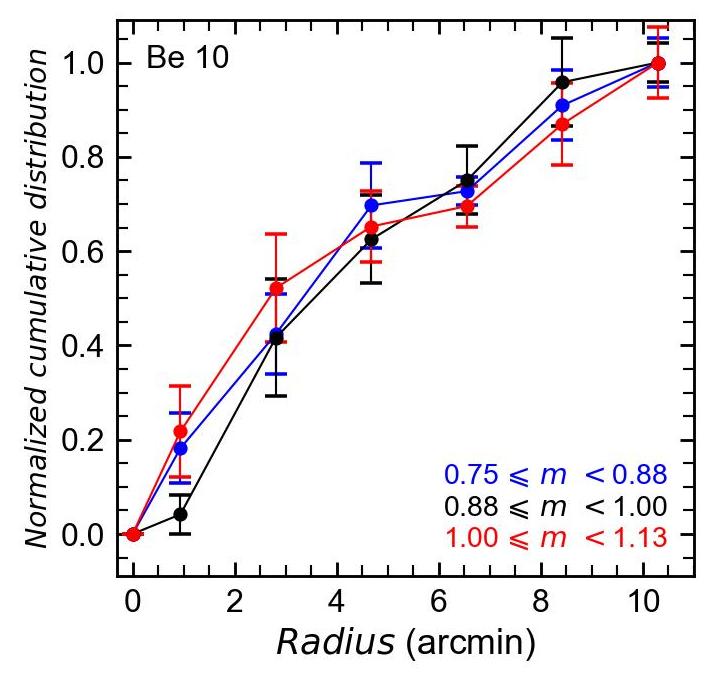}\hspace{5mm}
		\includegraphics[width=0.395\columnwidth]{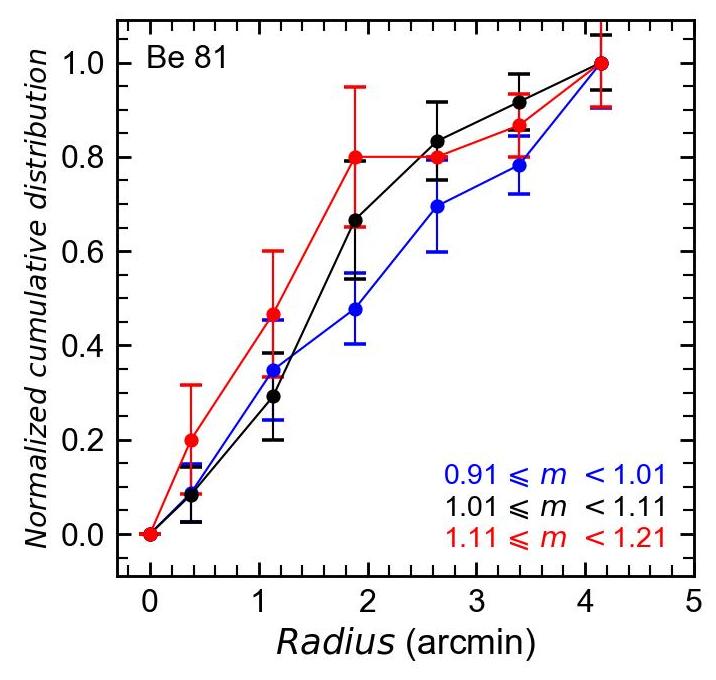}\vspace{3mm}
		\includegraphics[width=0.39\columnwidth]{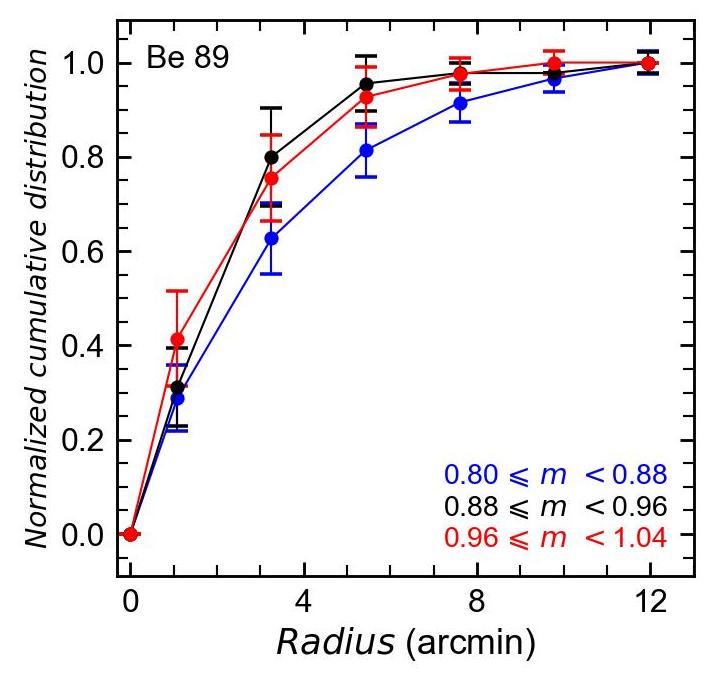}\hspace{5mm}
		\includegraphics[width=0.395\columnwidth]{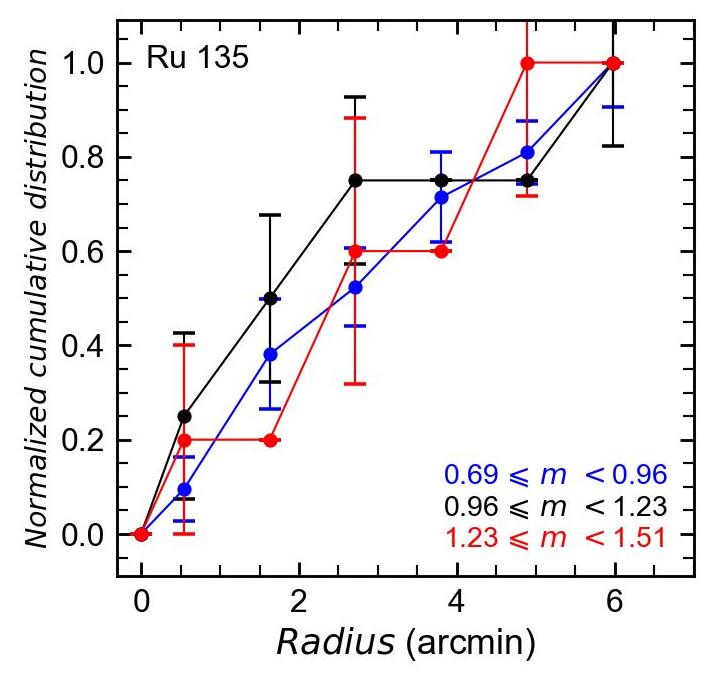} }
	\caption {The normalized cumulative distributions against cluster radius (arcmin) of four OCs in three different mass ranges ($m(m_{\odot})$). The vertical errors are from $\sqrt{N}$.}
\end{figure}

Some factors such as age, distance and selection of bright (high mass) stars are effective in these trends. As stated by \cite{Tarr2021},  old OCs are  mass segregated than young OCs. Old OCs with a significant number of stars show strong mass segregation ratio. These old OCs are on the average older than the OCs with few stars which show strongly mass segregation.  Cluster stars in the outer skirts of old OCs are on average lower than for young ones. In the sense  the proportion of stars in the outer parts of the OCs decreases towards old ages.  Besides this, bright/high mass stars are more likely to be selected than faint stars when distinguishing cluster members. In accordance with that, especially in the core region, there is a possibility that faint stars will not be selected as cluster members, due to the high density of stars. Therefore, there may be a bias in favor of bright stars.  Distance is a problem for observing faint stars, due to dense dust.

\section{Time-Scale Parameters and Tidal Radius}
The relaxation time, $t_{rlx}$  is defined as the time required for the stars in the core/halo of an open cluster to travel from one end of these regions to the other. Encounters between the cluster stars lead to an increased degree of energy equipartition. $t_{rlx}$ is obtained from the relation of \cite{Spitzer1971}
\begin{equation}
t_{rlx}=\frac{8.9\times10^5\sqrt{N}\times{R_{h}}^{3/2}}{log(0.4N)\times \sqrt{m}}
\end{equation}
where $N$, $R_{h}$, $m$ denote the cluster members (Col.~6 of Table 4), the radius containing half the cluster mass, the mean mass of the cluster members (Col.~5 of Table 4), respectively. Here  $R_{h}$ is estimated from the equation of \cite{Larsen2006},
\begin{equation}
R_{h}=0.547\times R_{c}\times(\frac{R_{t}}{R_{c}})^{0.486}
\end{equation}
where $R_{c}$ and $R_{t}$ are the core and tidal radii, respectively.  $R_{t}$ is from equation~(4). The dynamical evolutionary parameters of four OCs are estimated from
\begin{equation*}
	\tau=\frac{age}{t_{rlx}}
\end{equation*}
The ages are from Col.~7 of Table 2. The dissociation times of four OCs have been determined from the equation of \citet{Binney2008}, as given in the paper of \cite{Converse2011}.
\begin{equation}
t_{diss}=250~Myr(\frac{M}{300M_{\odot}})^{1/2}\times(\frac{R_{h}}{2~pc})^{-3/2}
\end{equation}
$M$ and $R_{h}$ are the total cluster mass and half mass radius of the clusters, respectively.
Here, tidal radius is the distance from the cluster center where gravitational acceleration caused by the cluster becomes equal to the tidal acceleration due to parent Galaxy \citep{vonHoerner1957}. As the cluster goes near its perigalacticon, a star at a large distance from the cluster center will be detached by galactic tidal forces whereas a star at a small distance will not. In other words, the tidal radius is linked to the stripping of stars from the cluster by the Galactic tidal field. Tidal interactions are essential to interprete the initial structure and dynamical evolution of the OCs \citep{Chumak2010}. 

Tidal radius $R_{t}$ is estimated from the equation of \cite{Kim2000}
\begin{equation}
R_{t}=(\frac{M} {2M_{G}})^{1/3}\times R_{G}	
\end{equation}
Here, $M$ is the total mass of the clusters (Col.4 of Table 4). The Galactic mass $M_{G}$ inside a galactocentric radius $R_{GC}$ (row~12 of Table 5) is obtained from the equation of \citet{Genzel1987},\\
\begin{equation}
M_{G}=2\times10^{8} M_{\odot} (\frac{R_{G}} {30 pc})^{1.2}	
\end{equation}
The tidal radius of a $10^{4} M_{\odot}$ cluster which is located at 30 pc from the globular cluster is $\sim$ 1 pc \citep{Kim2000}. $M_{G}$, $R_{t}$, $R_{h}$, and time scale parameters are listed in Cols.~2-7 of Table~6. 
\renewcommand{\tabcolsep}{5.4mm}
\renewcommand{\arraystretch}{1.2}
\begin{table}[!t]\label{Table-5}
	\begin{center}
		\caption{The weighted average radial velocities, ($V_R$) km s$^{-1}$ (row~1), space velocity components and rotational velocity  ($U$, $V$, $W$, $V_{\Phi}$) km s$^{-1}$ (rows~2-5), eccentricity (ecc) and peri and apogalactic distances, birth and present day distances (R$_{max}$, R$_{min}$, R$_{m}$, z$_{max}$, R$_{birth}$, R$_{GC}$) (kpc) (rows~6-12). The orbital angular momentum ($J_{z}$) (kpc km s$^{-1}$) (row~13). $T$ is the time of one revolution around the Galactic center as Myr (row~14). $N_{Rev}$ is the number of the revolution over the age of the cluster (last row).}\vspace{-1.5ex}
		\begin{tabular}{lcccc}
			\hline
			& Be\,10 & Be\,81 & Be\,89 & Ru\,135\\
			\hline
			$V_R$        & -36.7\hpm1.2 & 49.1\hpm0.9 & -44.1\hpm4.0 & -13.7\hpm1.4  \\ 
			$U$           &    37.90  &    56.59  &    29.66  &    -6.62 \\
			$V$           &   -11.61  &     3.74  &   -48.17  &    -8.98 \\
			$W$           &    -6.62  &     0.23  &     0.15  &   -11.76 \\
			$V_{\Phi}$    &  -243.83  &  -263.72  &  -206.08  &  -241.84 \\
			ecc           &     0.07  &     0.09  &     0.16  &     0.04 \\
			R$_{min}$     &     9.45  &     6.17  &     6.05  &     7.32 \\
			R$_{max}$     &    10.96  &     7.33  &     8.39  &     7.91 \\
			R$_m$         &    10.31  &     6.81  &     7.36  &     7.62 \\
			z$_{max}$     &     0.35  &     0.13  &     0.25  &     0.30 \\
			R$_{birth}$   &    10.08  &     6.55  &     6.80  &     7.66 \\
			R$_{GC}$      &     9.69  &     6.18  &     8.27  &     7.54 \\
			$J_z$         & -2365.21  & -1628.94  & -1704.42  & -1824.18 \\
			$T$           &    260    &     159   &    220    &    194 \\
			$N_{Rev}$     &    5.2    &     10.1  &     14.6  &    5.2 \\
			\hline
		\end{tabular}
	\end{center}
\end{table}\renewcommand{\tabcolsep}{1.2mm}
\renewcommand{\arraystretch}{1.2}
\begin{table*}[!h]\label{Table-6}
	\begin{center}
		\caption{Dynamical parameters. See Sect.~5 for their meanings.}\vspace{1ex}
		{\small
			\begin{tabular}{lAAAAAA}
				\hline
				Cluster      &\mcl{$M_G$}&\mcl{$R_t$}&\mcl{$R_h$}&\mcl{$t_{rlx}$}&\mcl{$t_{diss}$}&\mcl{$\tau$}\\
				&\mcl{$10^{9}\,M_{\odot}$}&\mcl{pc}&\mcl{pc}&\mcl{Myr}&\mcl{Myr}&\mcl{}\\
				\hline
				Be\,10  &  207.9&0.9  &  6.28&0.02  &  1.37&0.06  &  9.9&1.1  &  86.1&7.5  &  137&18 \\
				Be\,81  &  121.8&0.9  &  5.15&0.04  &  0.84&0.05  &  4.5&0.6  &  45.3&5.0  &  357&50 \\
				Be\,89  &  172.0&0.5  &  6.41&0.02  &  1.70&0.05  & 14.8&1.4  & 140.5&10.3 &  217&22 \\
				Ru\,135 &  149.6&0.5  &  5.14&0.02  &  0.60&0.02  &  2.8&0.4  &  23.4&2.4  &  363&60 \\
				\hline
			\end{tabular}
		}
	\end{center}
\end{table*}

\section{Kinematics and Orbital Parameters}
The Gaia radial velocities of the bright giants of Be~10 ($N= 11$), Be~81 ($N = 32$), Be~89 ($N = 24$), and Ru~135 ($N = 1$) have been taken from \cite{Soubiran2018,Soubiran2019}. Their weighted averages are listed in row~1 of Table 5.  Their heliocentric velocities ($U$, $V$, $W$) in the right-hand system have been obtained from the Gaia EDR3 radial velocities, the median proper motion components and the distances of these bright giants by using the algorithm of \cite{joh87}. Here the photometric distances (Col.~6 of Table~7 in it) are adopted. These space velocities have been transformed to the components $U'$, $V'$, $W'$ by correcting for the Solar motion $(U, V, W)_{\odot} = (+11.10, +12.24, +7.25)$ km s$^{-1}$ with respect to the local standard of rest (LSR) \cite{sch10}. Here, $R_{\odot}=8.2\pm0.1$ kpc \citep{bg16} and $V_{LSR}$ = 239\,km\,s$^{-1}$ \citep{bru11} are adopted. The heliocentric cartesian distances ($x'$, $y'$, $z'$) (kpc) and LRS-velocity components ($U'$, $V'$, $W'$) have been converted to Galactic Rest of Frame (GSR) i.e., ($x$, $y$, $z$) (kpc) and ($V_{x}$, $V_{y}$, $V_{z}$) from the equations of \cite{kep07}. The galactocentric velocity component ($V_{\Phi}$) (km s$^{-1}$) (or azimuthal velocity) in a cylindrical frame is estimated via 
\begin{equation*} 
	V_{\Phi} =  \frac{x V_{y} - y V_{x}}{R}
\end{equation*}
Here, $V_{\Phi}<0$ means prograde.  From the "MWPotential2014" code in the galpy-code library \footnote[1]{http://github.com/jobovy/galpy} written by \cite{bov15}, peri- and apo-galactic distances $(R_{min},~R_{max})$ (kpc) and the maximum height distance (z$_{max}$) (kpc) are obtained. The orbital eccentricity (ecc) is estimated via the relation
\begin{equation*} 
	ecc = \frac{R_{max}-R_{min}}{R_{max}+R_{min}}
\end{equation*}
Five evolved member's orbits have been integrated for the ages of four OCs (Col.~7 of Table 2) within the galactic potential. The galactic potential as sum of the galactic components is explained by \cite{bov15}. Their orbital angular momentum components  $J_{z}$ (kpc km s$^{-1}$) are calculated from the equation of \cite{kep07}. These all parameters are given in Table 5.
	
Rotational velocities $V_{\Phi}= 206-264$ km s$^{-1}$, the eccentricities, 0.04$-$0.16, and orbital angular momentum values (Table 5) indicate that four OCs have Galactic thin disk properties. x-y~(kpc) plane is known as projected on to the Galactic plane, whereas z-R (kpc) is the meridional plane. On x-y~(kpc) plane (top panels of Fig.~9), they follow a circular path around the Galactic center with the eccentricities, 0.04$-$0.16. From Fig.~9 (top panels), four OCs are orbiting near the Galactic disk, therefore they may be affected by the tidal forces of the disk. According to their revolution periods T~(Myr) around the Galactic center (row~14 of Table~5), Be~10 and Ru~135 have made nearly five revolutions around the center of the Galaxy.  For Be~81 and Be~89, their tours are nearly 10 to 15 (row~15 of Table~5). 
  
The orbits in z-R (kpc) plane (bottom panels of Fig.~9) have boxy-like type properties. Four OCs move in the meridional planes within the confined spaces. They are oscillating along z-axis.  Be 10's orbit is confined in a box of $\sim 9.5 < R_{gc} \leq 11.5$ kpc, therefore it is not interacting with the inner region of the Galaxy. The confined spaces are as the following, $\sim$$ 6.5 < R_{gc} \leq 8.0 $ kpc for Be 81,  $\sim 6.0 < R_{gc} \leq 8.7$ kpc for Be 89, $\sim 7.3 < R_{gc} \leq 7.8$ kpc for Ru 135. 

Their birth and present day positions in our Galaxy are shown with the filled blue and red dots , respectively (Fig.~9). These positions have been  calculated by using the "MWPotential2014" in the galpy-code package. For this, time is taken as zero (birth) and cluster age (present day). Their closest approaches to the Sun are  determined as $(d~(kpc),~t (Gyr))$ = (1.91, 1.35) (Be~10), (1.85, 0.16) (Be~81), (1.09, 0.09) (Be~89), (0.90, 1.00) (Ru~135), respectively.

\begin{figure*}[!t]\label{fig-9}
	\centering{
		\includegraphics[width=0.23\textwidth]{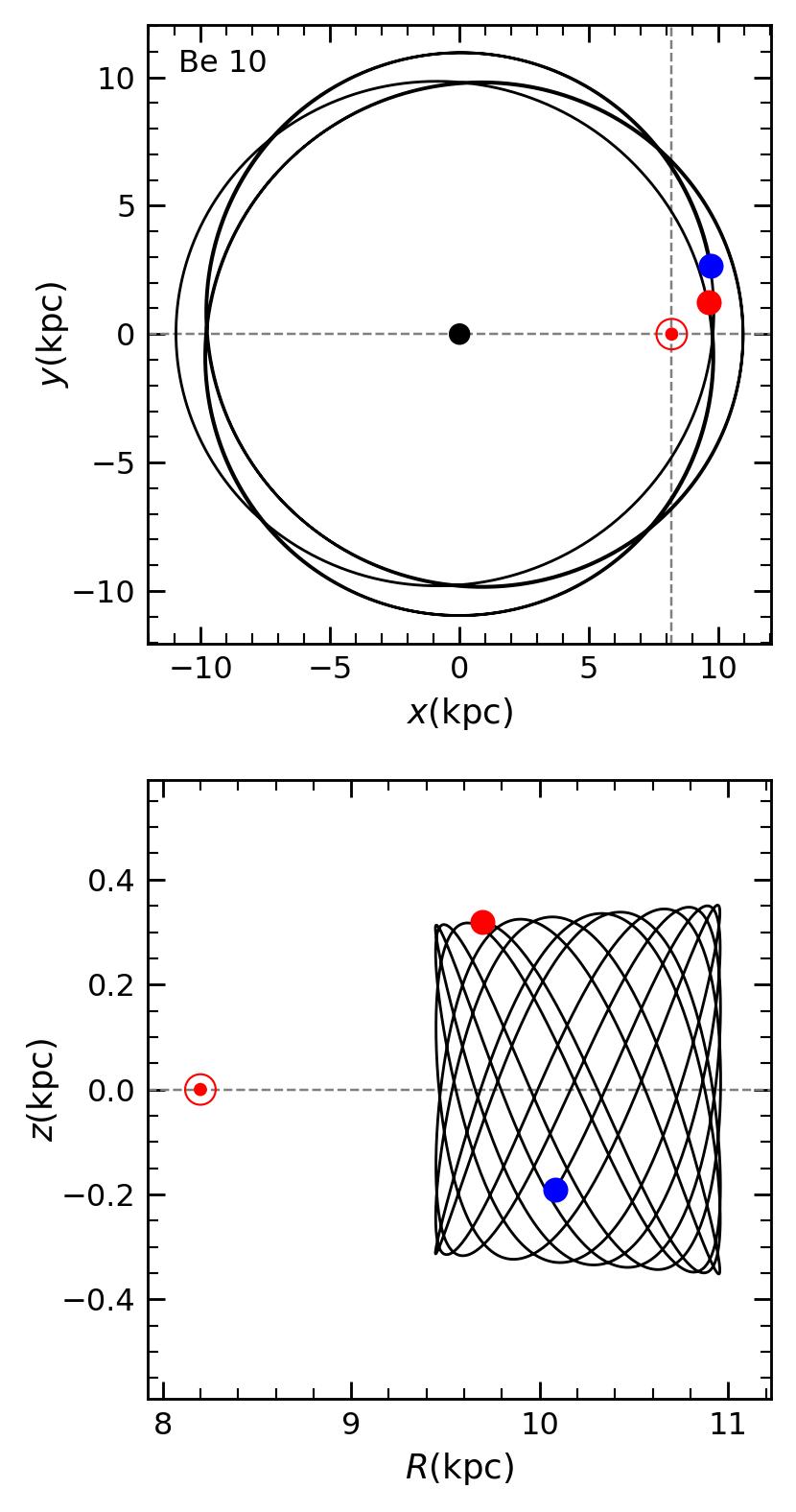}\hspace{0.5mm}
		\includegraphics[width=0.23\textwidth]{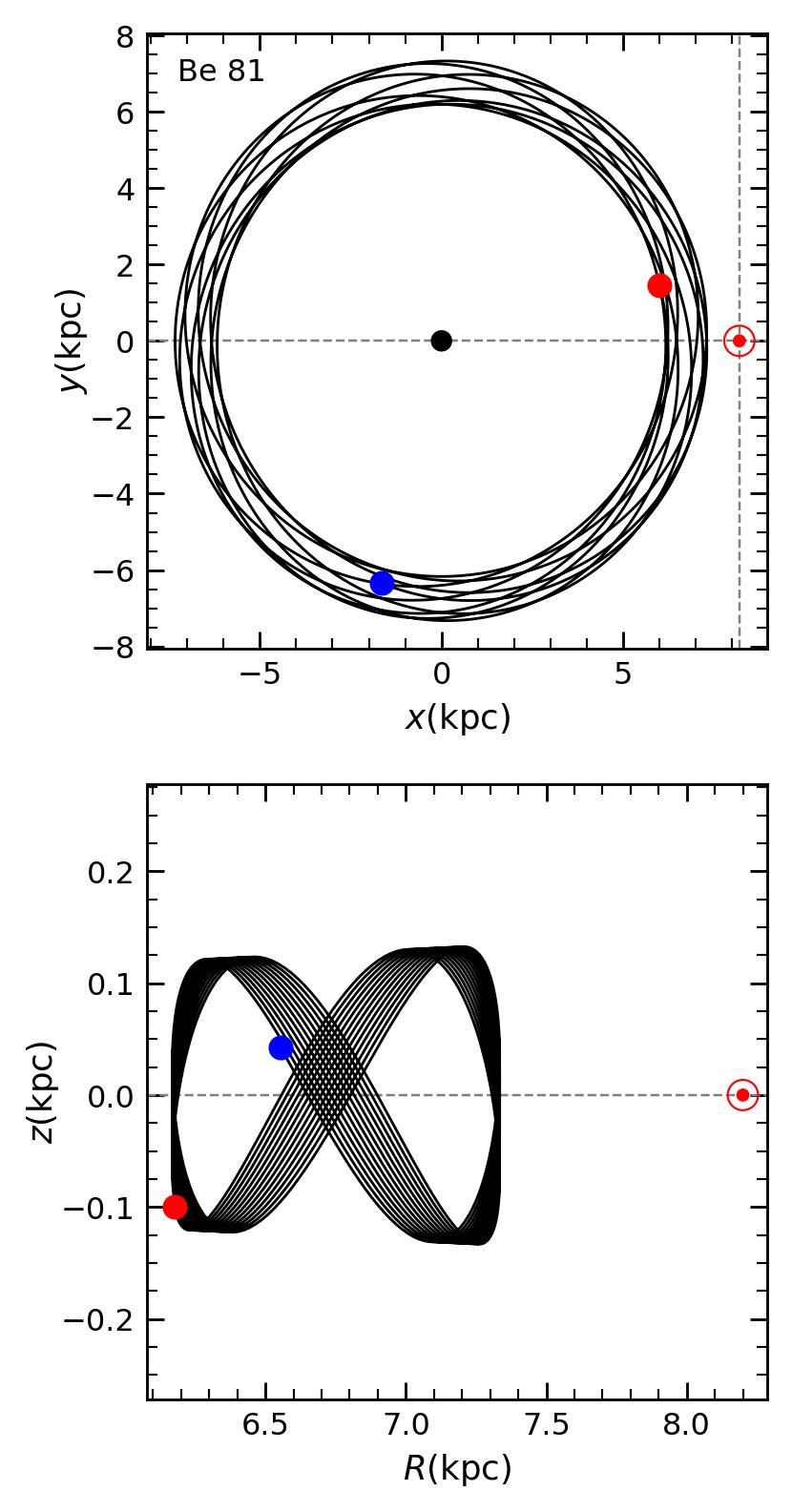}\vspace{0.5mm}
		\includegraphics[width=0.23\textwidth]{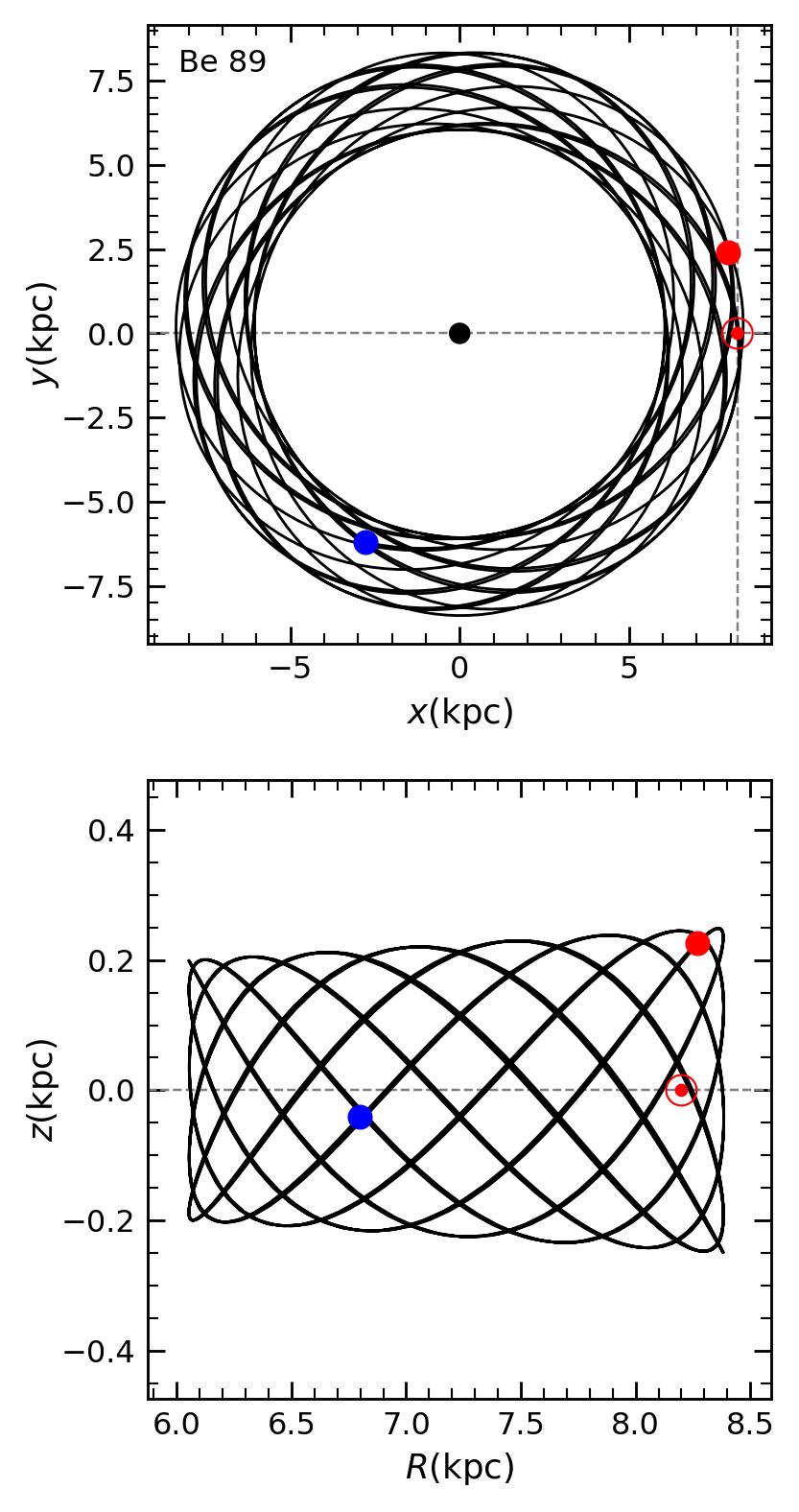}\hspace{0.5mm}
		\includegraphics[width=0.23\textwidth]{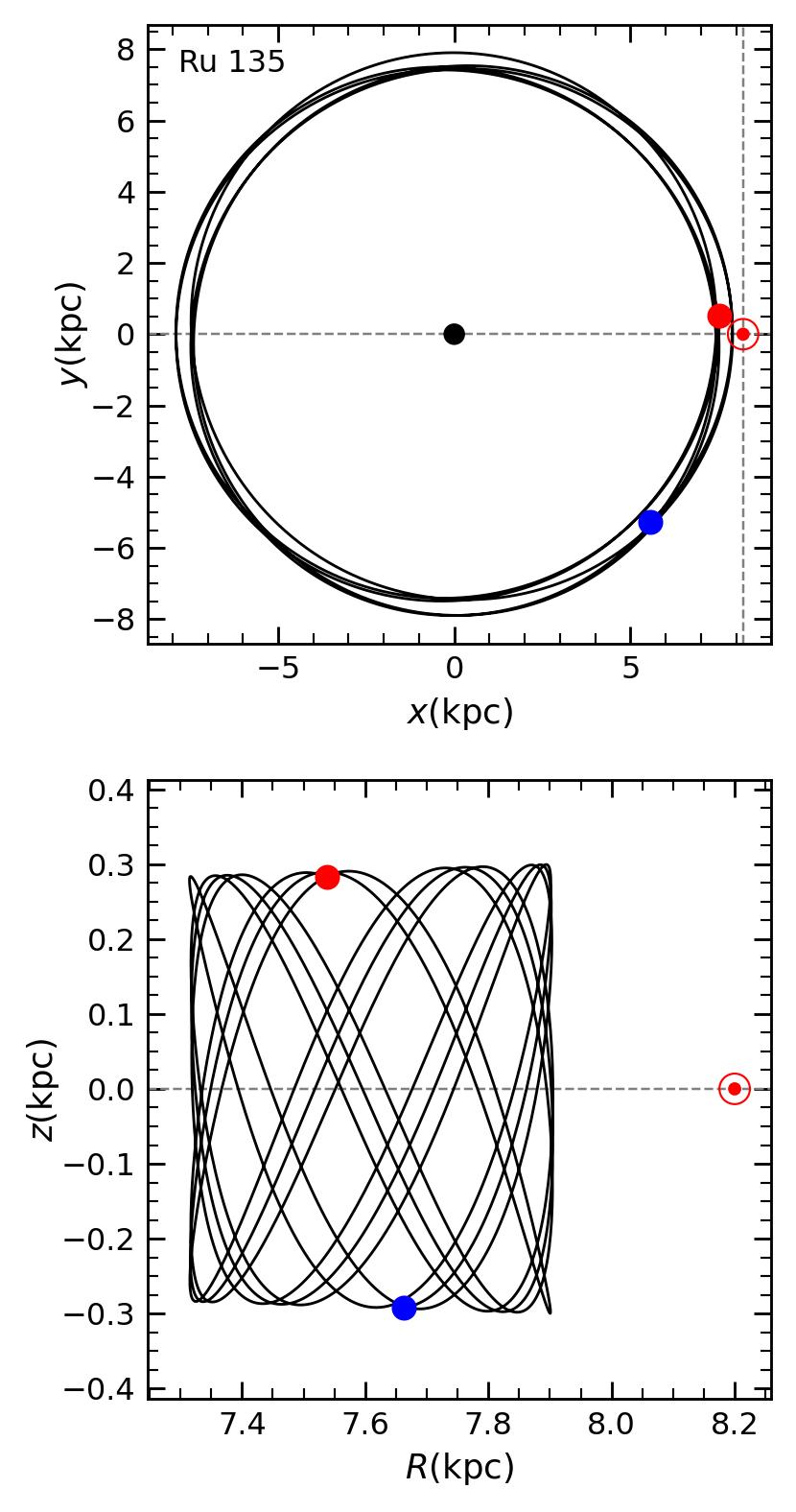}
	}
	\caption {Galactic orbits of four OCs. Trajectories represent the paths traveled by the OCs through their ages. The filled blue/red dots show their birth/present day positions. The red circle shows the Sun's location.}
\end{figure*}

\begin{figure}[b!]\label{fig-10}
	\centering{\includegraphics[width=0.60\columnwidth]{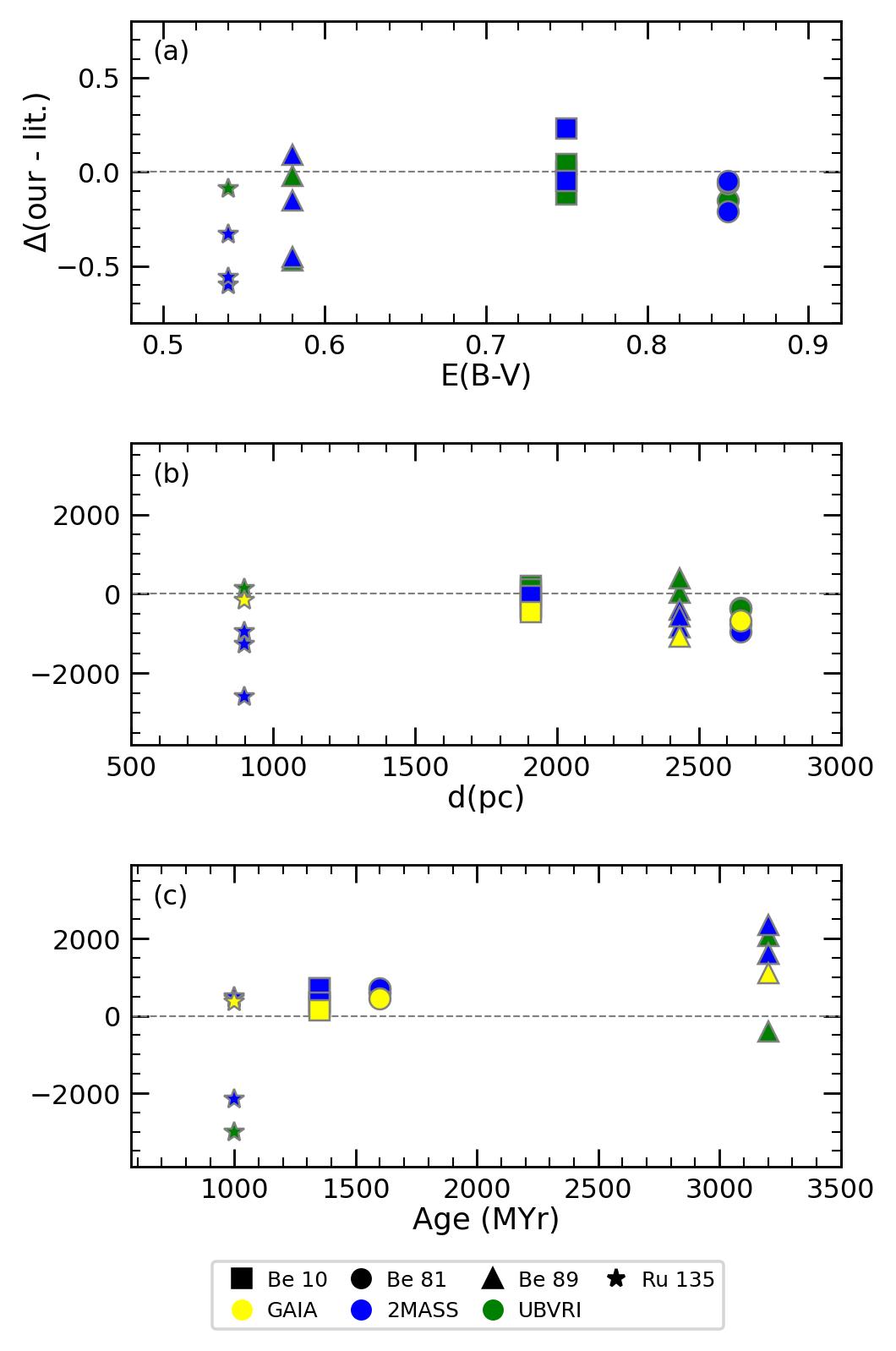}}
	\caption{The differences between this paper and literature for $E(B-V)$, $d~(pc)$, and $Age$~(Myr) in Table~2.}
\end{figure}

\section{Discussion and Conclusion}	
\subsection{Astrophysical parameters}	
The astrophysical values ($E(B-V)$, $d~(pc)$, and $Age$~(Myr)) together with the literature comparison are listed in Table~2. The comparisons are also displayed in Fig.~10. Reddening differences  are larger, up to 0.45$-$0.47 ($2MASS$/$CCD~UBVRI$) for Be~89 and 0.33$-$0.60 ($2MASS$) for Ru~135, respectively. Except this, our $E(B-V)$ values are in agreement with the literature within the discrepancies, $<$ 0.20 mag. The distance discrepancies to the literature are 61$-$456 pc (Be~10), 352$-$953 pc (Be~81), 33$-$1075 pc (Be 89), and 147$-$2600 pc (Ru~135), respectively.  Our distances are small, as compared to the ones of $2MASS$. There does not seem to be agreement even among the authors for $2MASS$/$CCD~UBVRI$ findings. Our distances for Be~89 and Ru~135 are larger but small for Be 81, as compared to the ones of the CCD-based $UBVRI$. For Be~10, the distance of \cite{lata2004} is larger than our value. The distances from the median parallaxes of four OCs (Table~1) are somewhat larger than the Gaia EDR3-isochrone fitting distances.

Note that the Gaia DR2 photometric distances of \cite{cantat2020} are larger than our distances. The discrepancy with the literature of the distance and age values of Be~81 is due to the fact that \cite{sag1998} and \cite{Donati2014} fit the isochrones at turn-off value, $V\sim 16.5$.  \cite{cantat2020} give a weight at turn-off point, $G\sim 15.53$ (dashed isochrone of right panel of Fig.~4), which is slightly fainter than our $G\sim 15.17$. 

From Table~2/Fig.~10(c), the ages of CCD-based UBVRI for four OCs are older than our ages.  However, for Be~89, our age is older than the one of \cite{Subramaniam2010}. 
The old ages of Ru~135 and Be~89 given by \cite{akkaya2010} are due to usage of the metal poor isochrones while fitting in their CMDs. Our ages of four OCs are older than the ones of $2MASS$, except for the old age value for Ru~135 of \cite{Kharchenko2013}.  From Fig.~10(c), the age differences (yellow symbols) between this paper  and the ones of \cite{cantat2020} are small at a level of 148$-$452 Myr for Be~10, Be~81 and Be~89.  For Ru~135, the discrepancy is somewhat larger, up to $\sim1000$ Myr.
The potential sources of discrepancies between our findings and the literature are of the adopted isochrones and the reddening values, the ground-based astrometric data, and different procedures for the selection of member stars, respectively, as stated by \cite{Paunzen2006} and \cite{Moitinho2010}.

\subsection{Dynamical Evolution}
Four OCs are in harmony with the increasing trend between $R_{core}$ and $R_{RDP}$ (Fig.~11a) \citep{Gunes2017, Camargo2009}. On average, larger cluster radii have larger core radii. Note that our OCs with old ages are prone to have small core radii, as shown by \cite{Tarr2021}. Here as an approximation, a value of $(R_{core},~R_{RDP})=(1.5,~7)$~pc of \cite{Gunes2017} is taken into account in classifying the cluster dimensions as small and large. 
\begin{figure}[!t]\label{fig-11}
	\centering{\includegraphics[width=0.60\columnwidth]{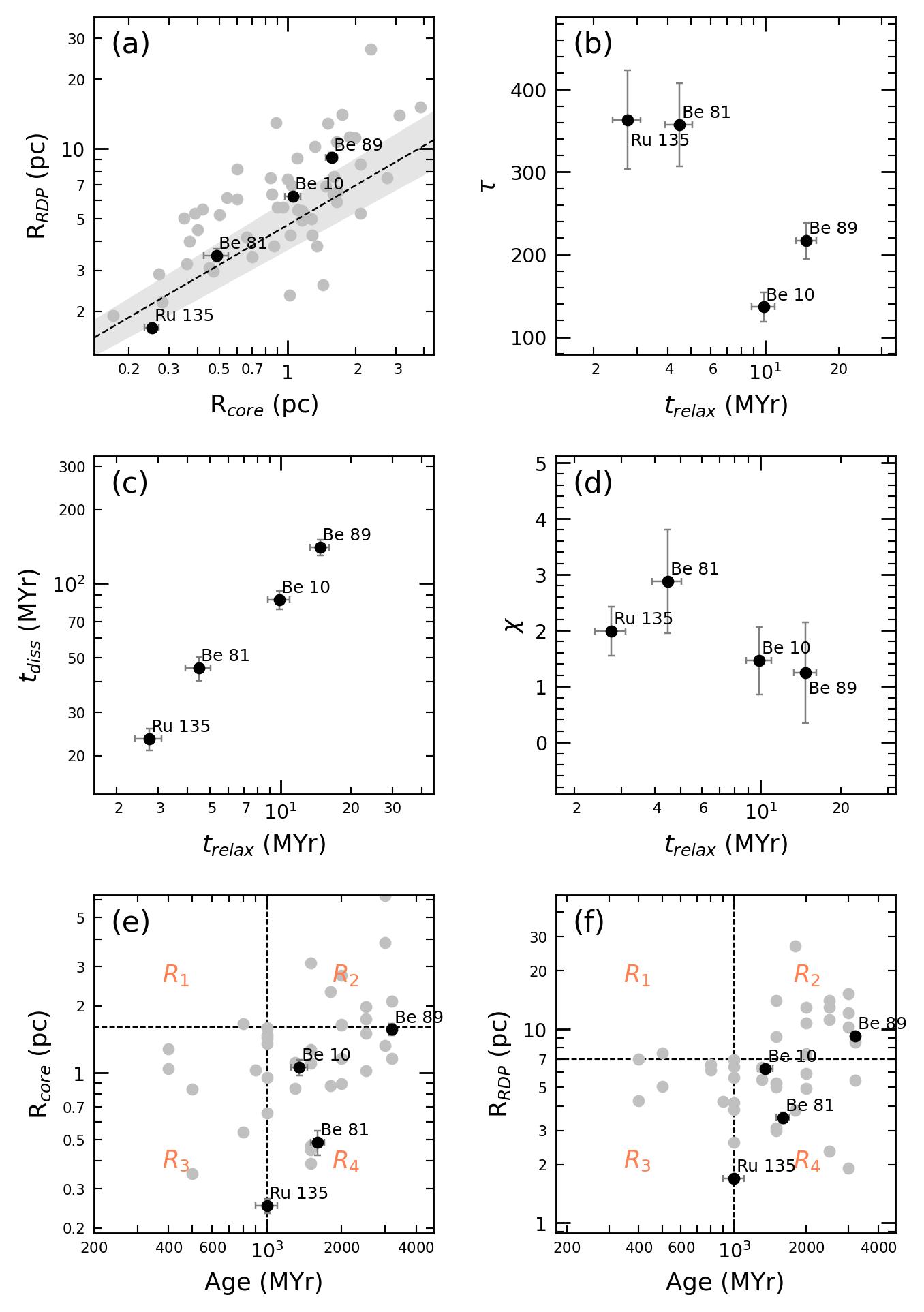}}
	\caption {$R_{RDP}$ versus $R_{core}$  (panel~a),  $\tau$ versus $t_{rlx}$ (panel~b), $t_{diss}$ versus $t_{rlx}$ (panel c), $\chi$ versus $t_{rlx}$ (panel~d),  $(R_{core},~R_{RDP})$ versus Age (Myr) (panels~e--f). The relation and its shaded area (1$\sigma$ uncertainty) in panel~(a), the labels $R1$-$R4$ and filled grey points of panels (e-f) are from \cite{Gunes2017}.}
\end{figure}

The dissociation times of four OCs are higher than their relaxation times (Fig.~11c/Col.~6 of Table~6).  This necessarily mean that four OCs disintegrate very early in the initial phase of quick expansion that accompanies the death of their most massive (bright) stars, as stated by \cite{Converse2011}. The relaxation times of four OCs are smaller than their ages (Col.~5 of Table~4 and Col.~7 of Table~2),  from this respective, they are dynamically relaxed. The steep overall MF slopes of four OCs (Fig.~11d) mean that their low mass stars outnumber their massive ones (Fig.~7 and Fig.~11d). Their large $\tau$ and relatively small $t_{rlx}$ values imply an advanced mass segregation. On account of this, their low mass stars are being lost to the field. 

Following the works of \cite{Camargo2009, Gunes2017, Cakmak2021}, four OCs are plotted on  $(R_{core},~R_{RDP})$ versus Age~(Myr) (panels~(e-f) of Fig.~11). This relationship is related to  survival and dissociation rates of clusters \citep{Camargo2009}. The meanings of the symbols are explained in caption of Fig.~11. Because Be~89 locates in $R2$ region (panels e$-$f), its core/outer parts show an expansion with time. However, Be~10 and Be~81 indicate the relatively shrinkage core/cluster radii ($R4$). Be~81 and Ru~135 locate in solar circle, and they are survived  although  they expose much to the external dynamical effects. Ru~135 (1.0~Gyr) with the small dimensions shrunk in size and mass with time, as it lost its star content because of the presence of massive GMCs, and tidal effects from disk and Bulge crossings as external perturbations, taking care its direction $(\ell=16^{\circ},~R_{GC}=7.5~kpc)$.  Instead of shrinking in size and mass with time, Ru~135 may have a primordial origin which may be related to high molecular gas density in Galactic directions \citep{Camargo2009, van1991}.
\begin{figure}[!b]\label{fig-12}
	\centering{\includegraphics[width=0.70\columnwidth]{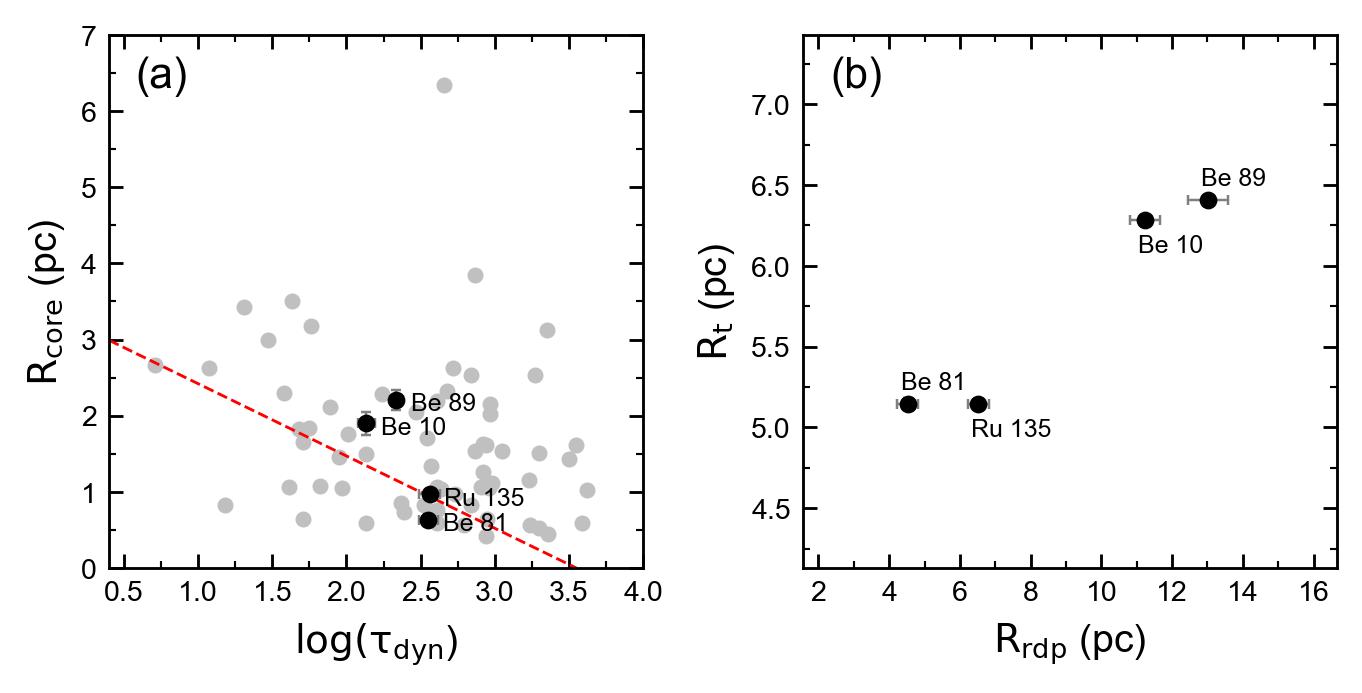}}
	\caption {$R_{core}$ versus $\log (\tau)$ (panel~a),  
		$R_{t}$ versus $R_{RDP}$ (panel~ b). The grey filled dots and the dashed line in panel (a) are from \cite{Angelo2021}.}
\end{figure}

For our sample OCs, the relations, $R_{core}$ versus $\log~(\tau)$ and $R_{t}$ versus $R_{RDP}$~(pc), respectively are presented in Fig.~12(a)$-$(b). Four OCs are in concordance with  the decreasing trend between $R_{core}$ and $\log (\tau)$ (panel~a). As stated by \citet{Angelo2020, Angelo2021}, this trend points out that our OCs with old ages are dynamically evolved, and they are losing their star contents to the field, due to internal/external dynamical processes. Be~89's tidal radius is less than its cluster radius (Fig.~12b/Tables~3 and 5). This means that its member stars lie within its tidal radius, in the sense it is gravitationally bound to the cluster.  For Be~10, Be~81 and Ru~135, their tidal radii are larger than their cluster radii. Thus, the cluster members beyond their tidal radii are gravitationally unbound to the cluster. The cluster members outside their tidal radii are more influenced by external potential of the Galaxy.

The expected masses of four OCs based on its peri-galacticon distance/its tidal radius are obtained from the relation \citep{King1962},
\begin{equation*} 
	M_{c}=3.5M_{G}\left(\frac{R_{t}}{R_{min}}\right)^3
\end{equation*}
where $R_{min}$ and  $M_{G}$ is the peri-galacticon distance and the Galaxy mass. $R_{t}$, $M_{G}$, and $R_{min}$, respectively, are taken from Tables~4$-$5. With these values, their expected masses  have been calculated as 214~$M_{\odot}$ (Be~10), 248~$M_{\odot}$ (Be~81), 716~$M_{\odot}$ (Be~89), and 181~$M_{\odot}$ (Ru~135), respectively. For instance, the expected mass 214~$M_{\odot}$ in Be\,10 decreases to $\sim110M_{\odot}$ (observed stellar total mass), because mass-loss processes such as evaporation and the external dynamical effects which have been acting on Be~10 for about 1.3~Gyr.

The orbits of Be~10 (Fig.~8) indicate that it passed a part of its time at galactosentric radius, $R_{GC}=9.69$ kpc, and then it was born at $R=10.08$ kpc, which would explain its solar abundance. According to the ($[M/H]$) versus R~(kpc) of \cite{lep11}, Be~10 occupies a region of $R > 9$ kpc (co-rotation gap at 9 kpc).  In the sense its origin may be attributed to different galactic radius or different star formation region \citep{lep11}.

\section{Acknowledgments}
We thank the anonymous referee for the comments and suggestions. We sepically thank to Charles Bonatto and Orhan G\"{u}ne\c{s} for the valuable comments. This paper has made use of results from the European Space Agency (ESA) space mission Gaia, the data from which were processed by the Gaia Data Processing and Analysis Consortium (DPAC). Funding for the DPAC has been provided by national institutions, in particular the institutions participating in the Gaia Multilateral Agreement. The Gaia mission website is http: //www.cosmos.esa.int/gaia. This paper has also made use of the WEBDA database, operated at the Institute for Astronomy of the University of Vienna. This publication also makes use of SIMBAD database-VizieR (http://vizier.u-strasbg.fr/viz-bin/VizieR?-source=II/246.).

\end{document}